\documentclass[
  pre,
  twocolumn,
  twoside,
  byrevtex,
  superscriptaddress,
  longbibliography,
  floatfix,
]{revtex4-1}

\usepackage[realmainfile]{currfile}
\newcommand{\suppmaterial}{Supplementary Material}

\usepackage[export]{adjustbox}

\usepackage{color}

\usepackage{graphicx,epsfig,verbatim,enumerate}
\usepackage{amssymb,amsmath}
\usepackage{ifthen}

\usepackage{longtable}

\usepackage{mathtools}

\newboolean{twocolswitch}

\newcommand{\sindex}[1]{}
\newcommand{\nindex}[1]{}

\newcommand{\www}[1]{\url{#1}}

\newcommand{\Req}[1]{Eq.~(\ref{#1})}

\usepackage{lettrine}

\usepackage{etoolbox,refcount}

\newcounter{countitems}
\newcounter{nextenumeratecount}
\newcommand{\setupcountitems}{%
  \stepcounter{nextenumeratecount}%
  \setcounter{countitems}{0}%
  \preto\item{\stepcounter{countitems}}%
}
\makeatletter
\newcommand{\computecountitems}{%
  \edef\@currentlabel{\number\c@countitems}%
  \label{countitems@\number\numexpr\value{nextenumeratecount}-1\relax}%
}
\newcommand{\nextenumeratecount}{%
  \getrefnumber{countitems@\number\c@nextenumeratecount}%
}
\makeatother
\AtBeginEnvironment{enumerate}{\setupcountitems}
\AtEndEnvironment{enumerate}{\computecountitems}

\usepackage{xfrac}

\newcommand{\onehalf}{\frac{1}{2}}

\newcommand{\flux}{\phi}

\newcommand{\fluxrankexponentlower}{\nu}

\newcommand{\zipfrank}{r}
\newcommand{\zipfexponent}{\zeta}

\newcommand{\zipfranktypesystema}[1]{\zipfrank_{\textnormal{#1},\indexaraw}}
\newcommand{\zipfranktypesystemb}[1]{\zipfrank_{\textnormal{#1},\indexbraw}}

\newcommand{\sizesymbol}{s}
\newcommand{\numbersymbol}{N}

\newcommand{\bigrank}{R}

\newcommand{\indexaraw}{1}
\newcommand{\indexbraw}{2}

\newcommand{\indexa}{(\indexaraw)}
\newcommand{\indexb}{(\indexbraw)}

\newcommand{\systemsymbol}{\Omega}
\newcommand{\elementsymbol}{\tau}

\newcommand{\systema}{\systemsymbol^{\indexa}}
\newcommand{\systemb}{\systemsymbol^{\indexb}}

\newcommand{\Ntypesa}{\numbersymbol_{\indexaraw}}
\newcommand{\Ntypesb}{\numbersymbol_{\indexbraw}}

\newcommand{\rtd}[1]{D^{\textnormal{R}}_{#1}}
\newcommand{\rtdelement}[1]{\delta D^{\textnormal{R}}_{#1,\elementsymbol}}

\newcommand{\rtdalpha}{\rtd{\alpha}}

\newcommand{\rtdalphavar}[1]{\rtd{#1}}

\newcommand{\rtdnorm}{\mathcal{N}_{\indexaraw,\indexbraw;\alpha}}
\newcommand{\invrtdnorm}{\frac{1}{\rtdnorm}}

\newcommand{\rtdnormalpha}[1]{\mathcal{N}_{\indexaraw,\indexbraw;#1}}
\newcommand{\invrtdnormalpha}[1]{\frac{1}{\rtdnormalpha{#1}}}

\newcommand{\rtdalphasystems}[2]{\rtdalpha(#1\,\|\,#2)}
\newcommand{\rtdalphasystemsOmega}{\rtdalphasystems{\systemsymbol_{\indexaraw}}{\systemsymbol_{\indexbraw}}}
\newcommand{\rtdalphasystemsRank}{\rtdalphasystems{\bigrank_{\indexaraw}}{\bigrank_{\indexbraw}}}

\newcommand{\rtdalphavarsystems}[3]{\rtdalphavar{#1}(#2\,\|\,#3)}

\newcommand{\rtdalphavarsystemsRank}[1]{\rtdalphavarsystems{#1}{\bigrank_{\indexaraw}}{\bigrank_{\indexbraw}}}

\newcommand{\rtdalphavarsystemsRankRand}[1]{\rtdalphavarsystems{#1; \textnormal{rand}}
   {\bigrank_{\indexaraw}}
   {\bigrank_{\indexbraw}}}

\newcommand{\bigrankordering}{\bigrank_{\indexaraw,\indexbraw;\alpha}}

\newcommand{\flipbooktwitter}{S1}
\newcommand{\flipbooktwitterRT}{S2}
\newcommand{\flipbooktwittertimediff}{S3}

\newcommand{\flipbooktrees}{S4}

\newcommand{\flipbookgirlsyears}{S5}
\newcommand{\flipbookboysyears}{S6}
\newcommand{\flipbookgirlsalphas}{S7}
\newcommand{\flipbookboysalphas}{S8}

\newcommand{\flipbookmarketcapsyears}{S9}

\newcommand{\flipbooktwittertrunc}{S10}
\newcommand{\flipbooktreestrunc}{S11}
\newcommand{\flipbookgirlnamestrunc}{S12}
\newcommand{\flipbookboynamestrunc}{S13}
\newcommand{\flipbookcompaniestrunc}{S14}

\newcommand{\flipbooknba}{S15}

\newcommand{\flipbookgoogleonegrams}{S16}
\newcommand{\flipbookgooglebigrams}{S17}
\newcommand{\flipbookgoogletrigrams}{S18}

\newcommand{\flipbookharrypotter}{S19}
\newcommand{\flipbookharrypotternocaps}{S20}

\newcommand{\flipbookdeathcauses}{S21}

\newcommand{\flipbookjobnames}{S22}

\setboolean{twocolswitch}{true}

\begin{document}

\title{\protect
Allotaxonometry and rank-turbulence divergence:\\
A universal instrument for comparing complex systems
}

\author{
  \firstname{Peter Sheridan}
  \surname{Dodds}
}

\email{peter.dodds@uvm.edu}

\affiliation{
  Computational Story Lab,
  Vermont Complex Systems Center,
  MassMutual Center of Excellence for Complex Systems and Data Science,
  Vermont Advanced Computing Core,
  University of Vermont,
  Burlington, VT 05401.
  }

\affiliation{
  Department of Mathematics \& Statistics,
  University of Vermont,
  Burlington, VT 05401.
  }

\author{
  \firstname{Joshua R.}
  \surname{Minot}
}

\affiliation{
  Computational Story Lab,
  Vermont Complex Systems Center,
  MassMutual Center of Excellence for Complex Systems and Data Science,
  Vermont Advanced Computing Core,
  University of Vermont,
  Burlington, VT 05401.
  }

\author{
  \firstname{Michael V.}
  \surname{Arnold}
}

\affiliation{
  Computational Story Lab,
  Vermont Complex Systems Center,
  MassMutual Center of Excellence for Complex Systems and Data Science,
  Vermont Advanced Computing Core,
  University of Vermont,
  Burlington, VT 05401.
  }

\author{
  \firstname{Thayer}
  \surname{Alshaabi}
}

\affiliation{
  Computational Story Lab,
  Vermont Complex Systems Center,
  MassMutual Center of Excellence for Complex Systems and Data Science,
  Vermont Advanced Computing Core,
  University of Vermont,
  Burlington, VT 05401.
  }

\author{
  \firstname{Jane Lydia}
  \surname{Adams}
}

\affiliation{
  Computational Story Lab,
  Vermont Complex Systems Center,
  MassMutual Center of Excellence for Complex Systems and Data Science,
  Vermont Advanced Computing Core,
  University of Vermont,
  Burlington, VT 05401.
  }

\author{
  \firstname{David Rushing}
  \surname{Dewhurst}
}

\affiliation{
  Computational Story Lab,
  Vermont Complex Systems Center,
  MassMutual Center of Excellence for Complex Systems and Data Science,
  Vermont Advanced Computing Core,
  University of Vermont,
  Burlington, VT 05401.
  }

\author{
  \firstname{Tyler J.}
  \surname{Gray}
}

\affiliation{
  Computational Story Lab,
  Vermont Complex Systems Center,
  MassMutual Center of Excellence for Complex Systems and Data Science,
  Vermont Advanced Computing Core,
  University of Vermont,
  Burlington, VT 05401.
  }

\affiliation{
  Department of Mathematics \& Statistics,
  University of Vermont,
  Burlington, VT 05401.
  }

\author{
\firstname{Morgan R.}
\surname{Frank}
}

\affiliation{
  Institute for Data, Systems, and Society,
  Massachusetts Institute of Technology,
  Cambridge,
  MA, 02139
}

\author{
  \firstname{Andrew J.}
  \surname{Reagan}
}

\affiliation{
  MassMutual Data Science,
  Amherst,
  MA 01002.
  }

\author{
  \firstname{Christopher M.}
  \surname{Danforth}
}

\affiliation{
  Computational Story Lab,
  Vermont Complex Systems Center,
  MassMutual Center of Excellence for Complex Systems and Data Science,
  Vermont Advanced Computing Core,
  University of Vermont,
  Burlington, VT 05401.
  }

\affiliation{
  Department of Mathematics \& Statistics,
  University of Vermont,
  Burlington, VT 05401.
  }

\date{\today}

\begin{abstract}
  \protect
  Complex systems often comprise
many kinds of components which vary over many orders of magnitude in size:
Populations of cities in countries,
individual and corporate wealth in economies,
species abundance in ecologies,
word frequency in natural language,
and node degree in complex networks.
Comparisons of component size distributions
for two complex systems---or a system with itself at
two different time points---generally
employ information-theoretic instruments,
such as Jensen-Shannon divergence.
We argue that these methods
lack transparency and adjustability,
and should not be applied when component probabilities
are non-sensible or are problematic to estimate.
Here, we introduce `allotaxonometry'
along with `rank-turbulence divergence',
a tunable instrument
for comparing any two (Zipfian) ranked lists of components.
We analytically develop our rank-based divergence in a series of steps,
and then establish a rank-based allotaxonograph which
pairs a map-like histogram for rank-rank pairs with an ordered list
of components according to divergence contribution.
We explore the performance of rank-turbulence divergence
for a series of distinct settings including:
Language use on Twitter and in books,
species abundance,
baby name popularity,
market capitalization,
performance in sports,
mortality causes,
and
job titles.
We provide a series of supplementary flipbooks
which demonstrate the
tunability and storytelling power of rank-based allotaxonometry.

\end{abstract}


\maketitle

\section{Introduction}
\label{sec:rankturbdiv.introduction}

\subsection{Instruments that capture complexity}
\label{subsec:rankturbdiv.introduction-instrument}

Science stands on the ability to describe and explain,
and precise quantification must ultimately secure any true understanding.
Description itself rests on well-defined, reproducible methods of measurement,
and over thousands of years, people have generated
many national museums' worth of physical and mathematical instruments
along with fundamental units of measurement.
Many instruments measure a single scale---in a plane's cockpit,
barometers, altimeters, and thermometers report pressure, height, and temperature.
And like a pilot flying a plane,
by using human-comprehendible dashboards of single-dimension instruments,
we are consequently able to successfully
monitor
and
manage
certain complex systems and processes.

But for complex phenomena made up of
a great many types of components
of greatly varying size---ecologies, stock markets, language---we
must confront two major problems with our dashboards of simple instruments~\cite{borland2019a}.

First, in the face of system scale, dashboards become overwhelming.
We find ourselves in high-dimensional, rapidly reconfiguring cockpits with instruments
constantly appearing and disappearing.
We need meters for every species, every company, every word.
As a consequence, we routinely reduce a system's description
to a few summary statistics, and often to only one~\cite{diamond1997a}.
We quantify the massive complexity of intellect
through intelligence quotients and grade point averages,
health through body mass index,
the complexity of civilizations by one number~\cite{turchin2018a},
and arguably anything by monetary value as an encoding of belief.
(Of course, for some systems, dimension reduction is possible
and we have essential techniques for doing so such as
as principal component analysis~\cite{strang2009a}.)
Relevant to our work here, information theoretic measures
such as Shannon's entropy or the Gini coefficient
are conspicuous single-number quantifications used across many fields,
whether or not there is any meaningful connection to the
optimal encoding of symbols for signal transmission~\cite{shannon1948a,jost2006a,shannon1956a}.

Second, enabling an  ability to discern change
is evidently an elemental feature of any scientific instrument.
Broken altimeters are a staple of stories where something goes wrong with a plane
(a plane-in-trouble the larger story trope unto itself).
While tracking changes in simple measures and statistics
is essential (the Dow Jones is up, today is warmer than yesterday),
the cognitive trap of the single number measurement means we 
miss seeing the internal dynamics, and this is especially true
when global statistics are constant.

To contend with scale and internal diversity of complex systems,
we need comprehendible, dynamically-adjusting dashboards.
For comparisons of complex systems,
we will argue for dynamic dashboards that have
two core elements~\cite{alajajian2017a}:
\begin{enumerate}
\item 
  A `big picture' map-like overview;
  and
\item
  A ranking of components afforded by
  a tunable measure that is as plain-spoken as possible.
\end{enumerate}

To help with our framing, we introduce a terminology family.
We will use `allotaxonomy' (\textit{other order})
to mean the general comparison of the structures of two complex systems;
`allotaxonometrics' to refer to quantified allotaxonomy;
and `allotaxonometers'
and `allotaxonographs'
for the instruments of allotaxonometrics.

\subsection{Zipf rankings, Zipf's law, and rank turbulence}
\label{subsec:rankturbdiv.introduction-rankings}

While the instrument we develop here will have broader application,
its construction focuses on two regular features of complex systems:
Heavy-tailed Zipf distributions (rather than laws),
and
what we will call `rank turbulence'---a phenomenon of system-system comparison.
We describe and discuss these two common signatures of complex systems in turn.

In general, we will consider systems where each component type $\elementsymbol$ has at
least one measurable---and hence rankable---``size'' $\sizesymbol_{\elementsymbol}$
where size
may be
count,
rate,
physical size,
monetary value,
scoring in sports by individual players,
and so on.
When a system's component types are ranked
in descending order of some size $\sizesymbol$,
we will write the size of the
$\zipfrank$th ranked component
as $\sizesymbol_{\zipfrank}$.
Though ranking is a widespread, everyday concept,
the associated language can be confusing:
High rank means low $\zipfrank$, and low rank means high $\zipfrank$.
The highest rank size is thus $\sizesymbol_{1}$.
(We accommodate tied ranks per Sec.~\ref{subsec:rankturbdiv.notation} below.)

Zipf's law is the specific observation that a Zipf ranking
obeys a  decaying power law~\cite{zipf1949a,simon1955a,newman2005b,coromina-murtra2010a}.
That is, the size $\sizesymbol_{\zipfrank}$
of the $\zipfrank$th ranked component
obeys the scaling
$\sizesymbol_{\zipfrank} \sim \zipfrank^{-\zipfexponent}$
where
the Zipf exponent
is $\zipfexponent > 0$.
The corresponding frequency distribution for component sizes will behave
as
$f(\sizesymbol) \sim \sizesymbol^{-\gamma}$
where
$\gamma = 1 + 1/\zipfexponent > 1$.

Power laws and their discontents aside, examples of heavy-tailed Zipf distributions abound,
with a few examples including
word and phrase frequency in language~\cite{gerlach2016a,williams2015a},
city populations~\cite{zipf1949a},
node degrees in scale-free networks~\cite{barabasi1999a},
firm size~\cite{axtell2001a},
and
numbers of dependencies for software packages~\cite{maillart2008a}.

We emphasize that our instrument is of use 
for comparing more general complex systems,
for which we need only a reasonably diverse set of component types,
and for which the Zipf ranking $\sizesymbol_{\zipfrank}$
may bear any kind of heavy-tailed distribution.
Below, we will explore systems with maximum component rank
between roughly $10^{2.5}$
and $10^{9}$.

There have been two persistent criticisms of Zipf's law,
one unfounded, the other true but misleading and central to our work here.
The first is that Zipf's law is a meaningless artifact
that arises for free through randomness~\cite{miller1957a,miller1965a};
this is negated by a simple analysis~\cite{ferrericancho2010a},
and moreover, theories of generative mechanisms
have long been elaborated and tested (and contested)
with the rich-get-richer mechanism proving to
be a pervasive underlying algorithm~\cite{simon1955a,mandelbrot1953a,maillart2008a,dodds2017a}.

The second enduring criticism is that Zipf's exponent
$\zipfexponent$ does not vary measurably,
whether it be over time for a given system or across comparable systems.
Zipf's law is often plotted with an unadorned rank $\zipfrank$
on the horizontal axis, but each rank represents a component type
from some vastly higher dimensional space of elements:
a language's lexicon, species in an ecology, corporations in an economy.

Thus, even if two meaningfully comparable systems
match exactly in a given Zipf ranking $\sizesymbol_{\zipfrank}$,
there may well be a rich variation in the ordering of components~\cite{gerlach2016a,pechenick2017a}.
With this understanding, in earlier work by our group on comparing Zipf rankings
of $n$-gram usage in large-scale texts,
we introduced
the concept of ``lexical turbulence''~\cite{pechenick2017a}.
We showed that in comparing word usage across decades in
the Google Books English Fiction (GBEF) corpus,
the flux of words across rank boundaries---rank flux $\flux_{\zipfrank}$---increased
as
$\flux_{\zipfrank} \sim \zipfrank^{\fluxrankexponentlower}$
(we found a break in scaling which we set aside here for simplicity~\cite{ferrericancho2001c,williams2015b}).
We observed superlinear scaling for
rank flux with $\fluxrankexponentlower > 1.2$:
Common words are relatively stable in rank, rare words much more unstable.

Here, we expand from the text-specific concept of lexical turbulence to a general one
of `rank turbulence', which in turn will help motivate our formulation
of a pragmatic `rank-turbulence divergence'.

\subsection{Motivation for a rank-based divergence}
\label{subsec:rankturbdiv.introduction-motivation}

In comparing complex systems, why should we use component size ranks rather
than probabilities or rates?
Indeed, there is a smorgasbord of ways to compare
two probability distributions for categorical data~\cite{deza2006a,cha2007a,cichocki2010a}.
Ref.~\cite{cha2007a} catalogs around 60 probability-based comparisons
which are variously distances, divergences,
similarities, fidelities, and inner products.
And Ref.~\cite{cichocki2010a} details three
sprawling, interrelated, single-parameter families of information-theoretic divergences.

Five main reasons push us away from probability-based divergences
and towards creating and using rank-based divergences.

First, normalization problems may arise from subsampling heavy-tailed
distributions~\cite{gerlach2016a,haegeman2013a}.
In natural ecological systems, for example, estimating the total number of
organisms is famously difficult~\cite{hill1973a,gotelli2011a,haegeman2013a,chao2014a}.
We can only then speak of relative rates and not absolute rates, and
even then only for common enough species.
For Twitter, subsampling 1-grams allows for robust estimation of the rates
of common 1-grams but not rare ones.

Second, not all component type characteristics can
be construed (or misconstrued) as probabilities or rates.
For example, rankings for many kinds of sports, at the team and player level and
not discounting the role of chance,
derive from scores achieved through repeated competition~\cite{merritt2014a,clauset2015a,kiley2016a}.

Third, in comparison with probability-based rankings,
we are able to more easily contend with components that appear
in only one of two systems under comparison.
We demonstrate this visualization feature as we build
rank-turbulence divergence (RTD) in the following sections.

Fourth, rank orderings
potentially allow for powerful and robust non-parametric statistical measures
such as Spearman's rank correlation coefficient.
All told,
while in moving to rankings
we may
trade information for some simplification,
we still preserve a great deal of meaningful structure.

Fifth and finally, rankings are an easily interpretable, ubiquitous construct.
Ranked lists suffuse media surrounding
entertainment (e.g., box office), music (Billboard charts), and sports.

The above notwithstanding,
distances based on comparisons of Zipf rankings
are to our knowledge relatively few,
focus on traditional comparative metrics
like
Kendall's Tau
and
Spearman's
rank correlation coefficient~\cite{fagin2003a},
and seem limited in application
to extremely small systems,
for example, comparing the top 20 to 50 ranked hits
from two different search engines~\cite{fagin2003a,bar-ilan2006a,webber2010a}.

And while we have argued for a rank-turbulence divergence here, we nevertheless
have separately constructed and explored a probability-turbulence divergence in~\cite{dodds2020g}.
Analogous in construction to rank-turbulence divergence,
we show that probability-turbulence divergence
is more sensitive to detailed system changes,
has distinct limiting behavior,
and corresponds to a suite of extant divergences.

\subsection{Paper outline}
\label{subsec:rankturbdiv.outline}

In Sec.~\ref{sec:rankturbdiv.rankturbdiv},
we develop rank-turbulence divergence by
(1) Establishing our notation and ranking process
(Sec.~\ref{subsec:rankturbdiv.notation});
(2) Creating and explaining a specific kind of rank-rank histogram
(Sec.~\ref{subsec:rankturbdiv.viz});
(3) Declaring a set of desired features for rank-turbulence divergence
(Sec.~\ref{subsec:rankturbdiv.features});
and then
(4) Building and refining a rank-turbulence divergence
that effectively captures these features
(Sec.~\ref{subsec:rankturbdiv.rankdiv}).

In Sec.~\ref{sec:rankturbdiv.graphicalinstrument},
we use all of these elements to realize
rank-turbulence divergence as a tunable instrument
for complex system comparison
through rank-turbulence divergence allotaxonographs.
To both support our general explanation
and explore systems in their own right,
we consider comparisons at different points in time for
four case studies:
1. daily word use on Twitter,
2. tree species abundance,
3. baby names in the US,
and
4. market capitalization for companies.

To help demonstrate the tunability of rank-turbulence divergence
and its behavior over time for dynamically evolving complex systems,
we provide Flipbooks of allotaxonographs as supplementary online material
on the arXiv and
at as part of the paper's online appendices:
\href{http://compstorylab.org/allotaxonometry/}{http://compstorylab.org/allotaxonometry/}.
Our Flipbooks expand on the paper's allotaxonomic analyses
to include
season point tallies for players in the National Basketball Association (NBA);
word usage in the Google Books corpus;
word usage in the seven Harry Potter books;
causes of death;
and
job advertisements.
As a guide, we outline all Flipbooks in Sec.~\ref{sec:rankturbdiv.flipbooks}.

We present details of datasets and code
in Sec.~\ref{sec:rankturbdiv.methods}, and
we round off our paper with some concluding thoughts
in Sec.~\ref{sec:rankturbdiv.concludingremarks}.

\section{Rank-Turbulence Divergence}
\label{sec:rankturbdiv.rankturbdiv}

\subsection{Notation, Ranking Methodology, and Exclusive Types}
\label{subsec:rankturbdiv.notation}

As mentioned in the introduction, we use Zipfian ranking~\cite{zipf1949a},
ordering a system $\systemsymbol$'s types from largest to smallest size
according to some measure (number, probability, mammalian fur density, etc.).
Again, we write $\sizesymbol_{\elementsymbol}$ for the size of
component type $\elementsymbol$.
We further indicate the rank of type $\elementsymbol$
as $\zipfrank_{\elementsymbol}$,
and the ordered set of all types and their ranks as $\bigrank_{\systemsymbol}$.

In the case of ties, we use the conventional 
tied rank method of fractional ranking.
For all types with the same size,
we assign the mean of the sequence of ranks
these types would occupy otherwise.
Retaining tied information in this way makes for
more sensible analytic treatment
(e.g., the sum of all ranks for $\numbersymbol$ types
will be
$
\frac{1}{2}
\numbersymbol(\numbersymbol+1)
$,
regardless of ties).
Ties (and near ties) will be important for
our visualizations of rank-turbulence divergence.

Given two systems,
$\systemsymbol_{\indexaraw}$
and
$\systemsymbol_{\indexbraw}$,
both comprised of component types
(e.g., the species of two ecosystems)
of
varying and rankable size (e.g., number of individuals in a species),
we express rank-turbulence divergence between these systems as
$\rtdalphasystemsOmega$.
In Sec.~\ref{subsec:rankturbdiv.rankdiv},
we will establish $\alpha$
as a single tunable parameter
with $0 \le \alpha < \infty$.

Whatever complexities these systems may contain---such as networks of components---we are
implicitly leaving them aside, but elaborations of our instrument will allow
their incorporation.
Thus to help with clarity, if we have two ranked lists to compare,
$\bigrank_{\indexaraw}$
and
$\bigrank_{\indexbraw}$,
we will more directly write
$\rtdalphasystemsRank$.

The divergences we will consider here
will all be expressible as linear sums
of per-type contributions, meaning we can write:
\begin{gather}
  \rtdalpha(
  \bigrank_{\indexaraw}
  \,\|\,
  \bigrank_{\indexbraw}
  )
  =
  \sum_{\elementsymbol \in \bigrankordering}
  \rtdelement{\alpha}(
  \bigrank_{\indexaraw}
  \,\|\,
  \bigrank_{\indexbraw}
  ).
  \label{eq:rankturbdiv.rankturbdiv_sum}
\end{gather}
We sort types by descending contribution,
$\rtdelement{\alpha}(
\bigrank_{\indexaraw}
\,\|\,
\bigrank_{\indexbraw}
)$,
indicating this ordering by the set $\bigrankordering$.

For the large-scale systems we are interested in, we expect
that the overlap of types between any two systems will be partial,
and generally far from complete.
Hashtags on Twitter for example are constantly being invented,
along with myriad lexical peculiarities
(keyboard mashings, misspelling, mistypings, and more~\cite{gray2019a}).

Therefore, when comparing two systems, we extend the list of types in both systems
to be the union of the types for both.
The sizes 
of types not present in a system
will be zero.
We will then naturally
assign the same equal last rank to all types that appear in one system
and not the other.

We call types that are present in one system only `exclusive types'.
When warranted, we will use expressions of the form
$\systema$-exclusive and $\systemb$-exclusive to indicate 
to which system an exclusive type belongs.

\subsection{Rank-Rank Histograms for Basic Allotaxonomy}
\label{subsec:rankturbdiv.viz}
    
\begin{figure*}
          \includegraphics[width=\textwidth]{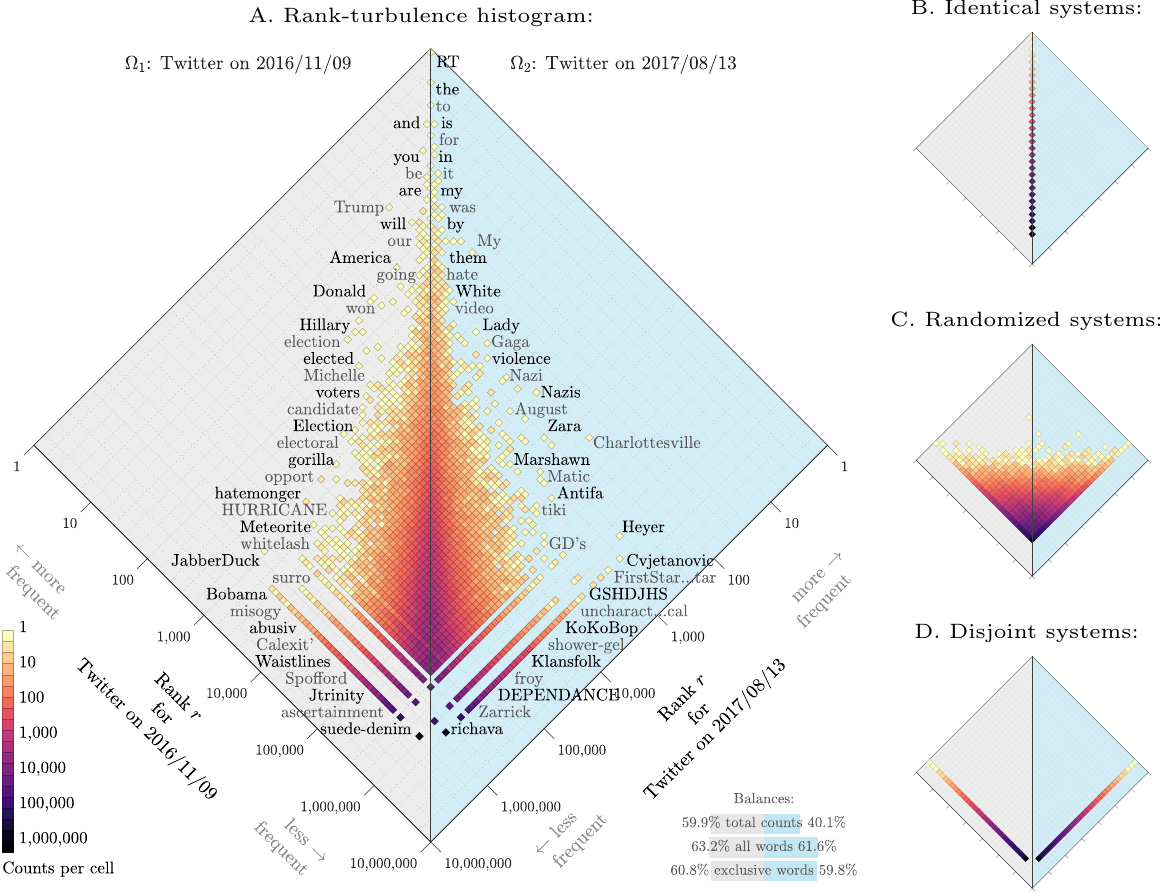}
  \caption{
    \textbf{A.}
    \textbf{
    An example allotaxonomic `rank-rank histogram'
    comparing word usage ranks on two days of Twitter, 2016/11/09 and 2017/08/13.}
    These dates are the day after the 2016 US presidential election
    and the day after the Charlottesville Unite the Right rally.
    Words are extracted first as 1-grams
    from tweets identified as English~\cite{alshaabi2020a}
    and then filtered to match simple latin characters
    (see Sec.~\ref{sec:rankturbdiv.methods.datasets}).
    We orient all histograms so that the comparison is left-right
    removing a potential misperception of causality.
    In general, we compare ranked lists of types for two systems
    $\systemsymbol_{\indexaraw}$
    and
    $\systemsymbol_{\indexbraw}$
    by first generating a merged list of types covering both systems.
    We then bin logarithmic rank-rank pairs
    $(\log_{10}\zipfrank_{\elementsymbol,\indexaraw},\log_{10}\zipfrank_{\elementsymbol,\indexbraw})$
    across all types and uniformly in logarithmic space.
    For bin counts, we use the perceptually uniform colormap magma~\cite{liu2018a},
    and place a scale in the bottom left corner.
    We automatically label words at the fringes of the histogram.
    Bins on either side of the central vertical line
    represent words that are used more often on the corresponding date.
    For example, `Charlottesville' was ranked 67,220 on 2016/11/09
    and 113 on 2017/08/13,
    while `Nazis' moved from $\zipfrank$=9,149 to 129.
    Words are given alternating shades of gray for improved readability.
    The discrete, separated lines of boxes nearest to each bottom axis comprise
    words that appear on Twitter on only that side's date: `exclusive types'.
    Moving up the histogram, the two distinct lines above the `exclusive-type lines'
    correspond to words that
    appear once and twice in the other system.
    The three horizontal bars in the lower right show system balances.
    The top bar indicates the balance of total counts of words for each day:
    59.9\% versus 40.1\%.
    The middle bar shows the percentage of the lexicon for the two days combined
    that appear on each day:
    63.2\% versus 61.6\%.
    And the bottom bar shows the percentage
    of words on each day that are exclusive:
    60.8\%
    and
    59.8\%.
    \textbf{B--D.}
    The three rank-rank histograms on the right
    show the special, benchmark cases of:
    \textbf{B.}
    A Zipf ranking for  compared with itself (vertical line; $\systemsymbol_{\indexaraw}$);
    \textbf{C.}
    A ranked list versus a random shuffling of component types
    ($\systemsymbol_{\indexaraw}$);
    and
    \textbf{D.}
    Two Zipf rankings for systems with no shared component types: a `vee' structure (we used $\systemsymbol_{\indexaraw}$ and $\systemsymbol_{\indexbraw}$, modifying words to prevent matches).
    For the cells in the main histograms in this paper, we use cell side lengths of 1/15 of an order of magnitude;
    we use 1/5 for plots \textbf{B--D}.
  }
  \label{fig:rankturbdiv.zipfturbulence-2016-11-09-2017-08-13-story-wrangler-all-rank-bare}
\end{figure*}

In Fig.~\ref{fig:rankturbdiv.zipfturbulence-2016-11-09-2017-08-13-story-wrangler-all-rank-bare}A,
we show an example of our base system-system comparison plot,
what we will call a `rank-rank histogram'.
We compare word usage on two days of Twitter:
The day after the 2016 US presidential election, 2016/11/09,
and the second day of the Charlottesville Unite the Right rally, 2017/08/13
(see Sec.~\ref{sec:rankturbdiv.methods.datasets} for description of datasets).
As we describe below, our histograms fully present the meaningful differences
between two Zipf distributions, allowing for divergence measures to be overlaid
in clarifying and easily interpretable ways (c.f.~\cite{monroe2008a,kessler2017a}).

To construct Fig.~\ref{fig:rankturbdiv.zipfturbulence-2016-11-09-2017-08-13-story-wrangler-all-rank-bare}A,
we first parse tweets into 1-grams (preserving case), find 1-gram frequencies for each day,
and then determine each day's separate ranked list of 1-grams according to those frequencies.
For both days, and purely by choice,
we take the subset of 1-grams that contain simple latin characters.
We next generate a merged list of simplified 1-grams observed on both days
and thereby obtain rank-rank pairs for all 1-grams.

For our histograms, we bin rank-rank pairs 
$(\zipfrank_{\elementsymbol,\indexaraw},\zipfrank_{\elementsymbol,\indexbraw})$
into cells uniformly in logarithmic space.
Cell width is adjustable; here we choose 1/15 of an order of magnitude.
We use a perceptually uniform colormap (magma~\cite{liu2018a}),
with the number of rank-rank pairs per cell increasing per
the lower left scale in
Fig.~\ref{fig:rankturbdiv.zipfturbulence-2016-11-09-2017-08-13-story-wrangler-all-rank-bare}A.
That the rank-rank pair counts per cell reach up towards $10^6$
should make clear that some form of histogram is necessary for attempting to visualize
the kind of rank turbulence we see here for Twitter.
A simple plot of all $(\zipfrank_{\elementsymbol,\indexaraw},\zipfrank_{\elementsymbol,\indexbraw})$ points
produces an incomprehensible density.

We orient our histograms in a diamond format,
rotating
the standard horizontal-vertical axes
$\pi/4$ counterclockwise.
We do so to eliminate a perceptual bias towards interpreting causality
(separately suggested in~\cite{bergstrom2018a}).
The vertical and horizontal coordinates in the rotated histogram
are proportional to
$\log_{10} \zipfrank_{\elementsymbol,\indexaraw} \zipfrank_{\elementsymbol,\indexbraw}$
(measured downwards)
and
$\log_{10} \zipfrank_{\elementsymbol,\indexbraw} / \zipfrank_{\elementsymbol,\indexaraw}$
(measured rightwards),
and these are dimensions we will encounter
later in our construction of rank-turbulence divergence.

Types that have higher rank in system
$\systemsymbol_{\indexaraw}$ will be represented by points on the left of the
vertical
$
\zipfrank_{\elementsymbol,\indexaraw}=\zipfrank_{\elementsymbol,\indexbraw}
$
line,
while
with have higher rank in system
$\systemsymbol_{\indexbraw}$ will appear on the right side.
Types falling along or near the center vertical line
have the same or similar ranks in both systems.

For all rank-rank histograms we show in the main paper,
we compare systems at different time points.
Time moving from left-to-right is a natural choice, and will govern our
arrangement of dynamically evolving systems.
In general however, comparisons between two systems may not involve any left-right
ordering, and the choice will be arbitrary,
(e.g., comparison of word usage in two books---see Flipbooks
described in Sec.~\ref{sec:rankturbdiv.flipbooks} for examples from the Harry Potter series---or
species abundance in two distinct ecological systems).

We automatically annotate words along the edges of the histogram.
To do so, we first specify a fixed bin size moving down the vertical axis.
For each bin and each side of the plot,
we find the word furthest away horizontally from the center line,
i.e., the word maximizing 
$| \log_{10} \zipfrank_{\elementsymbol,\indexaraw}/\zipfrank_{\elementsymbol,\indexbraw} |$.
Annotated words are oriented to the far side of
the point $(\zipfrank_{\elementsymbol,\indexaraw}, \zipfrank_{\elementsymbol,\indexbraw})$
relative to the center,
but are vertically centered by bin for overall clarity (meaning that their vertical
position relative to
$(\zipfrank_{\elementsymbol,\indexbraw}, \zipfrank_{\elementsymbol,\indexaraw})$
will fluctuate).
For these bare histograms with no divergence measure,
we also assign type names with alternating shades of gray for readability.
Where more than one word is equally far away from the center, we choose one
as a representative example.

To aid a user's perception of what meaning might
be rapidly conferred by a rank-rank histogram,
we highlight a selection of the annotated words
in Fig.~\ref{fig:rankturbdiv.zipfturbulence-2016-11-09-2017-08-13-story-wrangler-all-rank-bare}A.
Broadly, there are four main regions:
1. The top of the diamond;
2. The sides of the histogram;
3. The lower linear and point structures of the histogram;
and
4. The bottom of the diamond.

Types appearing towards the top of the diamond rank high for both systems.
For Fig.~\ref{fig:rankturbdiv.zipfturbulence-2016-11-09-2017-08-13-story-wrangler-all-rank-bare}A,
the 1-gram `RT' is the most common word on both days:
$\zipfranktypesystema{RT}
=
\zipfranktypesystemb{RT}
=
1$.
Signifying retweet, `RT' is an important---if Twitter-specific---functional structure,
indicating the strength of echoing on Twitter.
The words `the' and `to' are ranked 2nd and 3rd on both dates,
while `and' and `is' are ranked 4th and 4th on
2016/11/09 and reversed to 5th and 4th on 2017/08/13,
leading to their offset locations.
Such changes of high rank types will be important in analyzing many kinds of systems,
and we will see later that they are only picked up by certain divergences.

Moving down the histogram, we see that turbulence starts to become noticeable
around $\zipfrank=10^2$,
and we see increasingly less common and differentiating words appear.
Types appearing furthest horizontally from the center vertical axis show
the most relative change in rank.
On 2016/11/09, `Trump' stands out relative to nearby words.
Further down, `America, `Donald', `voted', and `election' are
all clearly off-axis.
On 2017/08/13,
the words `Charlottesville' and `Heyer' are most prominent
(Heather Heyer was a protester
who was murdered by vehicular homicide on August 12, 2017).

While 2016 election and Charlottesville terms dominate the sides of the histogram,
unrelated names and events also appear.
On the left of the histogram and/or list, we see the `Harambe' and `gorilla' 
and `Meteorite' while
on the right, we find Lady Gaga and Zara Larsson (both performed concerts),
and the Korean band BTS
which was enjoying its rise to ultrafame over this time period~\cite{dodds2019a}.
Harambe was a gorilla who was killed in a Cincinnati zoo after a
boy entered his enclosure in 2016/05.
Harambe became part of various internet memes
including ones putting him forward as a  write-in candidate for US president.

The separated lines and points at the bottom of the histogram
arise from logarithmic spacing.
For systems with heavy-tailed Zipf distributions for discrete sizes,
we often observe many types of the least size.
Here, where type size is word count, we have many hapax legomena---words
that appear only once in a corpus.
For books approximately obeying Zipf's law,
the fraction of a lexicon that appears is around 1/2~\cite{simon1955a}---the rare are legion.

Moving upwards from the bottom, the three separated lines 
in Fig.~\ref{fig:rankturbdiv.zipfturbulence-2016-11-09-2017-08-13-story-wrangler-all-rank-bare}A's histogram
correspond to words appearing zero times, once, and twice
on the other side's day.
We define `exclusive types' as those types that zero times in the other system,
i.e., those types that appear along the bottom separated lines of the histogram.

For example, at the extreme of the lowest line on the right,
we see `Cvjetanovic', a $\systemb$-exclusive word that is highly ranked on 2017/08/13
($\zipfranktypesystemb{Cvjetanovic}$=672).
The word is the last name of a member of Identity Evropa who was part
of the Unite the Right Rally; a photo of him holding a tiki torch and yelling
was widely circulated~\cite{wikipedia-identity-evropa2019a}.
The word `Cvjetanovic'
did not appear on 2016/11/09 and with zero counts,
is tied with many other words that only appear on 2017/08/13
($\zipfranktypesystema{Cvjetanovic}$=1,552,865).
As another example, the word `Heyer' appeared once on 2016/11/09 and is consequently
part of the second discrete line on the right side.

The least important and least differentiating types appear at the
bottom of the histogram.
These types are low rank in both systems.
The bottommost annotations in
Fig.~\ref{fig:rankturbdiv.zipfturbulence-2016-11-09-2017-08-13-story-wrangler-all-rank-bare}A,
`suede-denim' and `richava'
appear once on the dates of their respective sides.
These creatures of the lexical abyss are just two examples
of on the order of $10^6$ words
appearing once on only one of the two dates
(see the count scale in the lower left of
Fig.~\ref{fig:rankturbdiv.zipfturbulence-2016-11-09-2017-08-13-story-wrangler-all-rank-bare}A).

We emphasize that types annotated at or near the bottom of the diamond cannot be important
individually---no divergence measure should present `richava'
as a meaningful word in itself for these two days of Twitter.
Even so, indicating a few examples these of rare and unimportant words
along the bottom of the histogram
provides a helpful check that this is indeed the case.
With the aim of improving the instrument's
affordance of understanding,
when we introduce rank-turbulence divergence,
we will fade annotations according
to type-level divergence contributions.
Annotations for doubly rare types will always be strongly backgrounded.

Fig.~\ref{fig:rankturbdiv.zipfturbulence-2016-11-09-2017-08-13-story-wrangler-all-rank-bare}B--D
show examples of three extremes of how systems might compare on rank-rank histograms.
For real-world data, we will we will see various imprints of these three limiting cases.

In Fig.~\ref{fig:rankturbdiv.zipfturbulence-2016-11-09-2017-08-13-story-wrangler-all-rank-bare}B,
we compare the Zipf ranking for identical systems
($\systemsymbol_{\indexaraw}$ from
Fig.~\ref{fig:rankturbdiv.zipfturbulence-2016-11-09-2017-08-13-story-wrangler-all-rank-bare}A).
The outcome is a colormap version of the system's Zipf distribution
arranged on the vertical
$
\zipfrank_{\elementsymbol,\indexaraw}=\zipfrank_{\elementsymbol,\indexbraw}
$
line.

In Fig.~\ref{fig:rankturbdiv.zipfturbulence-2016-11-09-2017-08-13-story-wrangler-all-rank-bare}C,
we present the visualization of a system compared with a randomized version of
itself.
The nature of logarithms means that the lower triangle is well filled with
density growing with increasing rank.
Using a linear scale, we would see a statistically uniform histogram.

Finally,
in Fig.~\ref{fig:rankturbdiv.zipfturbulence-2016-11-09-2017-08-13-story-wrangler-all-rank-bare}D,
we compare Zipf distributions for systems with completely distinct sets of types.
After merging types across systems, ranking of types for each system places
all types of the other system in a tie for last place.
The result is two marginal Zipf distributions forming a `vee'.
We have already seen examples of these linear features
in Fig.~\ref{fig:rankturbdiv.zipfturbulence-2016-11-09-2017-08-13-story-wrangler-all-rank-bare}A.
If system component lists are sufficiently truncated---whether by measurement limitations or
by choice---we will also see these kinds of marginal structures appear
but in an inconsistent fashion.
We will discuss truncation effects further in
Sec.~\ref{subsec:rankturbdiv.truncation},
after introducing rank-turbulence divergence.

\subsection{Desirable Allotaxonometric Features for Rank-Turbulence Divergence}
\label{subsec:rankturbdiv.features}

On their own, our annotated rank-rank histograms
give a map-like overview
of how two systems differ.
For Twitter,
Fig.~\ref{fig:rankturbdiv.zipfturbulence-2016-11-09-2017-08-13-story-wrangler-all-rank-bare}A
presents a clear texture of words associated with the 2016 US election on the left
and the 2017 events of Charlottesville on the right.
But which words are most important?
How do we compare the relatively rare `Heyer' with the common `My', both words
that have higher ranks on 2017/08/13?

Our goal now is to construct a rank-based divergence for comparing complex systems,
and to function as an instrument overlaying rank-rank histograms.
We would like our divergence to be able to bear
the following \nextenumeratecount\ descriptors, which range from concrete and simple
to qualitative:
\begin{enumerate}
\item
  Rank-based: Directly built for comparing ranked lists generated by any meaningful ordering.
\item
  Symmetric:
  $
  \rtdalpha(
  \bigrank_{\indexaraw}
  \,\|\,
  \bigrank_{\indexbraw}
  )
  =
  \rtdalpha(
  \bigrank_{\indexbraw}
  \,\|\,
  \bigrank_{\indexaraw}
  ).
  $
\item
  Semi-positive:
  $
  \rtdalpha(
  \bigrank_{\indexaraw}
  \,\|\,
  \bigrank_{\indexbraw}
  )
  \ge 0$,
  and
  $
  \rtdalpha(
  \bigrank_{\indexaraw}
  \,\|\,
  \bigrank_{\indexbraw}
  ) = 0$
    only if
  the systems are formed by the same components with matching rankings,
  $
  \bigrank_{\indexaraw}
  =
  \bigrank_{\indexbraw}.
  $
\item
  Metric-capable:
  Given the preceding two conditions are met, we would need $\rtdalpha$ to also satisfy the triangle inequality.
\item
  Scale and unit invariant:
  This is automatic because rankings will not change if either one or both systems are
  rescaled in their entirety, or remeasured according to a different system of units.
\item
  Linearly separable, for interpretability.
  As framed in
  \Req{eq:rankturbdiv.rankturbdiv_sum},
  each type
  $\elementsymbol$
  additively contributes 
  to rank-turbulence divergence
  a quantity
  $
  \rtdelement{\alpha}
  (
  \bigrank_{\indexaraw}
  \,\|\,
  \bigrank_{\indexbraw}
  )$,
  allowing for simple ranking of types to assess importance.
\item
  Subsystem applicable: Ranked lists of any principled subset may be equally well compared (e.g., hashtags on Twitter, stock prices of a certain sector, etc.).
\item
  Effective across system sizes, possibly size independent:
  While not being explicitly interpretable as certain probability divergences
  (e.g., Kullback-Leibler divergence),
  rank-turbulence divergence
  $\rtdalphasystemsOmega$
  should be normalizable to allow for sensible comparisons of rank-turbulence divergences
  across system sizes.
  Linear separability means that whatever normalization we use,
  the ordering of contributions of individual types will be unchanged.
\item
  Zipfophilic: 
  Rank-turbulence divergence should be applicable to systems
  with rank-ordered component size distributions
  that are heavy-tailed.
\item
  Tunable:
  The acknowledgment that while many stand-alone divergences exist
  for probability distributions~\cite{cha2007a,cichocki2010a},
  in practice there are families of divergences on offer,
  and these have the potential to be adaptive and
  provide much more power and insight~\cite{cichocki2010a}.
\item
  Storyfinding: Features 1--\number\numexpr\value{enumi}-1\relax\
  will ideally combine
  to help us rapidly see which types are most important
  in distinguishing two ranked lists.
\end{enumerate}

\subsection{Development of Rank-Turbulence Divergence}
\label{subsec:rankturbdiv.rankdiv}

With these features in mind, we move now to properly constructing
our conception of rank-turbulence divergence.
We begin with the observation that by definition,
a type $\elementsymbol$'s Zipfian rank is inversely related
to its size.
We thus will want to deal with inverses of ranks.

Given element $\elementsymbol$ has a Zipfian rank
$\zipfrank_{\elementsymbol,\indexaraw}$
in system $\indexaraw$
and
$\zipfrank_{\elementsymbol,\indexbraw}$
in system $\indexbraw$,
a raw starting point for an element-level divergence
incorporating rank inverses would be:
\begin{equation}
  \centering
  \left\lvert
  \frac{1}{\zipfrank_{\elementsymbol,\indexaraw}}
  -
  \frac{1}{\zipfrank_{\elementsymbol,\indexbraw}}
  \right\rvert.
  \label{eq:rankturbdiv.elementranksimple}
\end{equation}
As we will demonstrate later,
experimentation with this fixed form reveals a bias towards
types with high ranks (again, the highest rank is $\zipfrank$=1).

We modify the above expression by introducing a
parameter $\alpha$:
\begin{equation}
  \left\lvert
  \frac{1}{\left[\zipfrank_{\elementsymbol,\indexaraw}\right]^{\alpha}}
  -
  \frac{1}{\left[\zipfrank_{\elementsymbol,\indexbraw}\right]^{\alpha}}
  \right\rvert^{1/\alpha}.
  \label{eq:rankturbdiv.elementrankturbdiv}
\end{equation}
We now have tunability:
As $\alpha \rightarrow 0$,
high ranked types are increasingly dampened
relative to low ranked ones.
For words in texts, for example, the weight of common words and rare
words will become increasingly closer together.
(Our construction and its behavior are
in parts resemblant of but distinct from
that of generalized entropy~\cite{renyi1961a,tsallis2001a,keylock2005a}
and 
Hill numbers in ecology~\cite{hill1973a,jost2006a}.)

At the other end of the dial, $\alpha \rightarrow \infty$,
high rank types will dominate.
For texts, function words will prevail while
the contributions of rare words will vanish.

The $\alpha \rightarrow \infty$ limit
will prove to be a natural parameter endpoint for
rank-turbulence divergence when we realize it as an instrument,
and is something we wish to preserve as we address the
$\alpha \rightarrow 0$ limit.

However, the limit of $\alpha \rightarrow 0$
in \Req{eq:rankturbdiv.elementrankturbdiv}
does not yet behave as
we might hope.
We see that if $\zipfrank_{\elementsymbol,\indexaraw} \ne \zipfrank_{\elementsymbol,\indexbraw}$,
\Req{eq:rankturbdiv.elementrankturbdiv} tends towards
\begin{equation}
  \alpha^{1/\alpha}
  \left\lvert
  \ln
  \frac{\zipfrank_{\elementsymbol,\indexaraw}
  }{\zipfrank_{\elementsymbol,\indexbraw}}
  \right\rvert^{1/\alpha},
  \label{eq:rankturbdiv.elementrankturbdivfail}
\end{equation}
which in turn will tend toward $\infty$ as $\alpha \rightarrow 0$.

In considering how to remedy this problematic limit,
we observe that \Req{eq:rankturbdiv.elementrankturbdivfail}
contains a readily interpretable structure which we
have already encountered in the preceding section:
the log-ratio of ranks.
In Sec.~\ref{subsec:rankturbdiv.viz}, 
we established a graphical interpretation
for the rank-rank histogram in
Fig.~\ref{fig:rankturbdiv.zipfturbulence-2016-11-09-2017-08-13-story-wrangler-all-rank-bare}A.
We identify
$
\left\lvert
\ln
\frac{\zipfrank_{\elementsymbol,\indexaraw}
}{\zipfrank_{\elementsymbol,\indexbraw}}
\right\rvert
=
\left\lvert
\ln
\zipfrank_{\elementsymbol,\indexaraw}
-
\ln
\zipfrank_{\elementsymbol,\indexbraw}
\right\rvert
$
as being proportional to the horizontal distance from the
$(\log_{10}\zipfrank_{\elementsymbol,\indexaraw},\log_{10}\zipfrank_{\elementsymbol,\indexbraw})$
point
to the 
vertical midline.

To preserve the core of
\Req{eq:rankturbdiv.elementrankturbdiv},
$
\left\lvert
\frac{1}{\left[\zipfrank_{\elementsymbol,\indexaraw}\right]^{\alpha}}
-
\frac{1}{\left[\zipfrank_{\elementsymbol,\indexbraw}\right]^{\alpha}}
\right\rvert^{1/\alpha},
$
maintain the form of the large $\alpha$ limit,
fashion a well-behaved
$\alpha \rightarrow 0$ limit,
and to only use modifications that are monotonic in $\alpha$,
we introduce a prefactor and adjust the exponent in
\Req{eq:rankturbdiv.elementrankturbdiv} as follows:
\begin{equation}
  \frac{\alpha+1}{\alpha}
  \left\lvert
  \frac{1}{\left[\zipfrank_{\elementsymbol,\indexaraw}\right]^{\alpha}}
  -
  \frac{1}{\left[\zipfrank_{\elementsymbol,\indexbraw}\right]^{\alpha}}
  \right\rvert^{1/(\alpha+1)}.
  \label{eq:rankturbdiv.elementrankturbdiv_good}
\end{equation}
The $\alpha \rightarrow 0$ limit is now simply
$
\left\lvert
\ln
\frac{\zipfrank_{\elementsymbol,\indexaraw}
}{\zipfrank_{\elementsymbol,\indexbraw}}
\right\rvert,
$
while the 
$\alpha \rightarrow \infty$
limit is unchanged.
(We note that an alternate modification of simply
introducing a prefactor of $\alpha^{-1/\alpha}$
to \Req{eq:rankturbdiv.elementrankturbdiv}
fails
the requirement of monotonicity.)

Finally, in summing over all types and 
incorporating a normalization prefactor $\rtdnorm$,
we have our prototype, single-parameter rank-turbulence divergence:
\begin{align}
  &
  \rtdalpha(
  \bigrank_{\indexaraw}
  \,\|\,
  \bigrank_{\indexbraw}
  )
  =
  \sum_{\elementsymbol \in \bigrankordering}
  \rtdelement{\alpha}(
  \bigrank_{\indexaraw}
  \,\|\,
  \bigrank_{\indexbraw}
  )
  \nonumber
  \\
  &
  =
  \invrtdnorm
  \frac{\alpha+1}{\alpha}
  \sum_{\elementsymbol \in \bigrankordering}
  \left\lvert
  \frac{1}{\left[\zipfrank_{\elementsymbol,\indexaraw}\right]^{\alpha}}
  -
  \frac{1}{\left[\zipfrank_{\elementsymbol,\indexbraw}\right]^{\alpha}}
  \right\rvert^{1/(\alpha+1)}.
  \label{eq:rankturbdiv.rankturbdiv_good}
\end{align}

While analytic forms for the normalization factor $\rtdnorm$ could be constructed,
we take a numerical approach.
We compute $\rtdnorm$ by taking the two systems to be disjoint while
maintaining their underlying Zipf distributions.
Thus, we ensure
$
0
\le 
\rtdalpha(
\bigrank_{\indexaraw}
\,\|\,
\bigrank_{\indexbraw}
)
\le
1 
$
where the limits of 0 and 1
correspond, respectively,
to the two systems having
identical 
and disjoint
Zipf distributions.

To determine $\rtdnorm$,
we observe that if the Zipf distributions are disjoint,
then in $\systema$'s merged ranking
the rank of all $\systemb$ types will
be
$\zipfrank = \Ntypesa + \onehalf\Ntypesb$,
where
$\Ntypesa$
and
$\Ntypesb$
are the number of distinct types in each system.
Similarly, $\systemb$'s merged ranking will
have all of $\systema$'s types in last place
with rank $\zipfrank = \Ntypesb + \onehalf\Ntypesa$.
The normalization is then:
\begin{align}
  \rtdnorm
  &
  =
  \frac{\alpha+1}{\alpha}
  \sum_{\elementsymbol \in \bigrank_{\indexaraw}}
  \left\lvert
  \frac{1}{\left[\zipfrank_{\elementsymbol,\indexaraw}\right]^{\alpha}}
  -
  \frac{1}{\left[\Ntypesa + \onehalf\Ntypesb\right]^{\alpha}}
  \right\rvert^{1/(\alpha+1)}
  \nonumber
  \\
  &
  +
  \frac{\alpha+1}{\alpha}
  \sum_{\elementsymbol \in \bigrank_{\indexaraw}}
  \left\lvert
  \frac{1}{\left[\Ntypesb + \onehalf\Ntypesa\right]^{\alpha}}
  -
  \frac{1}{\left[\zipfrank_{\elementsymbol,\indexbraw}\right]^{\alpha}}
  \right\rvert^{1/(\alpha+1)}.
  \label{eq:rankturbdiv.rankturbdiv_norm}
\end{align}

We note that for a disjoint pair of systems, their randomized versions
will necessarily still be disjoint,
and
$\rtdalphavarsystemsRankRand{\alpha} = 1$.

\subsection{Tunability of Rank-Turbulence Divergence: Limits}
\label{sec:rankturbdiv.tunability}

We will use rank-turbulence divergence's tunability to accentuate
more rare ($\alpha \rightarrow 0$)
or more common types ($\alpha \rightarrow \infty$).
For reference, we lay out the full expressions for these two limits,
and will later see their graphical realizations.
Per our construction of \Req{eq:rankturbdiv.rankturbdiv_good},
in the limit of $\alpha \rightarrow 0$,
we have 
\begin{equation}
  \rtd{0}(
  \bigrank_{\indexaraw}
  \,\|\,
  \bigrank_{\indexbraw}
  )
  =
  \sum_{\elementsymbol \in \bigrankordering}
  \rtdelement{0}
  =
  \invrtdnormalpha{0}
  \sum_{\elementsymbol \in \bigrankordering}
  \left\lvert
  \ln
  \frac{\zipfrank_{\elementsymbol,\indexaraw}
  }{\zipfrank_{\elementsymbol,\indexbraw}}
  \right\rvert,
  \label{eq:rankturbdiv.elementrankturbdiv_zerolimit}
\end{equation}
where
\begin{align}
  \rtdnormalpha{0}
  &
  =
  \sum_{\elementsymbol \in \bigrank_{\indexaraw}}
  \left\lvert
  \ln
  \frac{\zipfrank_{\elementsymbol,\indexaraw}
  }{\Ntypesa + \onehalf\Ntypesb}
  \right\rvert
  +
  \sum_{\elementsymbol \in \bigrank_{\indexbraw}}
   \left\lvert
  \ln
  \frac{\zipfrank_{\elementsymbol,\indexbraw}
  }{\onehalf\Ntypesa + \Ntypesb}
  \right\rvert.
  \label{eq:rankturbdiv.elementrankturbdiv_zerolimit_norm}
\end{align}
Types experiencing the largest relative change in rank will
feature most strongly, and these
are types that are rare in one system, and extremely common in the other.
Because of the term
$
\ln
\frac{\zipfrank_{\elementsymbol,\indexaraw}
}{\zipfrank_{\elementsymbol,\indexbraw}},
$
the $\alpha=0$ limit for rank-turbulence divergence
is most resemblant of
the Kullback–Leibler and Jeffrey divergences~\cite{deza2006a}.

In the limit of $\alpha \rightarrow \infty$, we have instead
\begin{align}
  &
  \rtd{\infty}(
  \bigrank_{\indexaraw}
  \,\|\,
  \bigrank_{\indexbraw}
  )
  =
  \sum_{\elementsymbol \in \bigrankordering}
  \rtdelement{\infty}
  \nonumber
  \\
  &
  =
  \invrtdnormalpha{\infty}
  \sum_{\elementsymbol \in \bigrankordering}
  \left(
  1
  -
  \delta_{
    \zipfrank_{\elementsymbol,\indexaraw}
    \zipfrank_{\elementsymbol,\indexbraw}
    }
  \right)
  \max_{\elementsymbol}
  \left\{
  \frac{
    1
  }{
    \zipfrank_{\elementsymbol,\indexaraw}
  },
  \frac{
    1
  }{
    \zipfrank_{\elementsymbol,\indexbraw}
  }
  \right\}.
  \label{eq:rankturbdiv.elementrankturbdiv_infinitylimit}
\end{align}
Having the lowest values of $1/\zipfrank$,
highest-rank types will
dominate the $\alpha \rightarrow \infty$ limit.
The normalization factor for $\alpha=\infty$ is:
\begin{gather}
  \rtdnormalpha{\infty}
  =
  \sum_{\elementsymbol \in \bigrank_{\indexaraw}}
  \frac{
    1
  }{
    \zipfrank_{\elementsymbol,\indexaraw}
  }
  +
  \sum_{\elementsymbol \in \bigrank_{\indexbraw}}
  \frac{
    1
  }{
    \zipfrank_{\elementsymbol,\indexbraw}
  }.
  \label{eq:rankturbdiv.elementrankturbdiv_infinitylimit_norm}
\end{gather}
For probability-based divergences,
the $\alpha=\infty$ limit
for rank-turbulence divergence
aligns with the Motyka distance~\cite{deza2006a,cha2007a}.

Because we are interested in real, finite systems,
we are not concerned with convergence.
Nevertheless, with appropriate treatment,
infinite theoretical systems could
be evaluated.

\section{Rank-Turbulence Divergence Graphs as Allotaxonometric Instruments}
\label{sec:rankturbdiv.graphicalinstrument}

\subsection{Anatomy of an allotaxonograph with word usage on Twitter as an example}
\label{subsec:rankturbdiv.construction}

\begin{figure*}
  \includegraphics[width=1.1\textwidth,center]{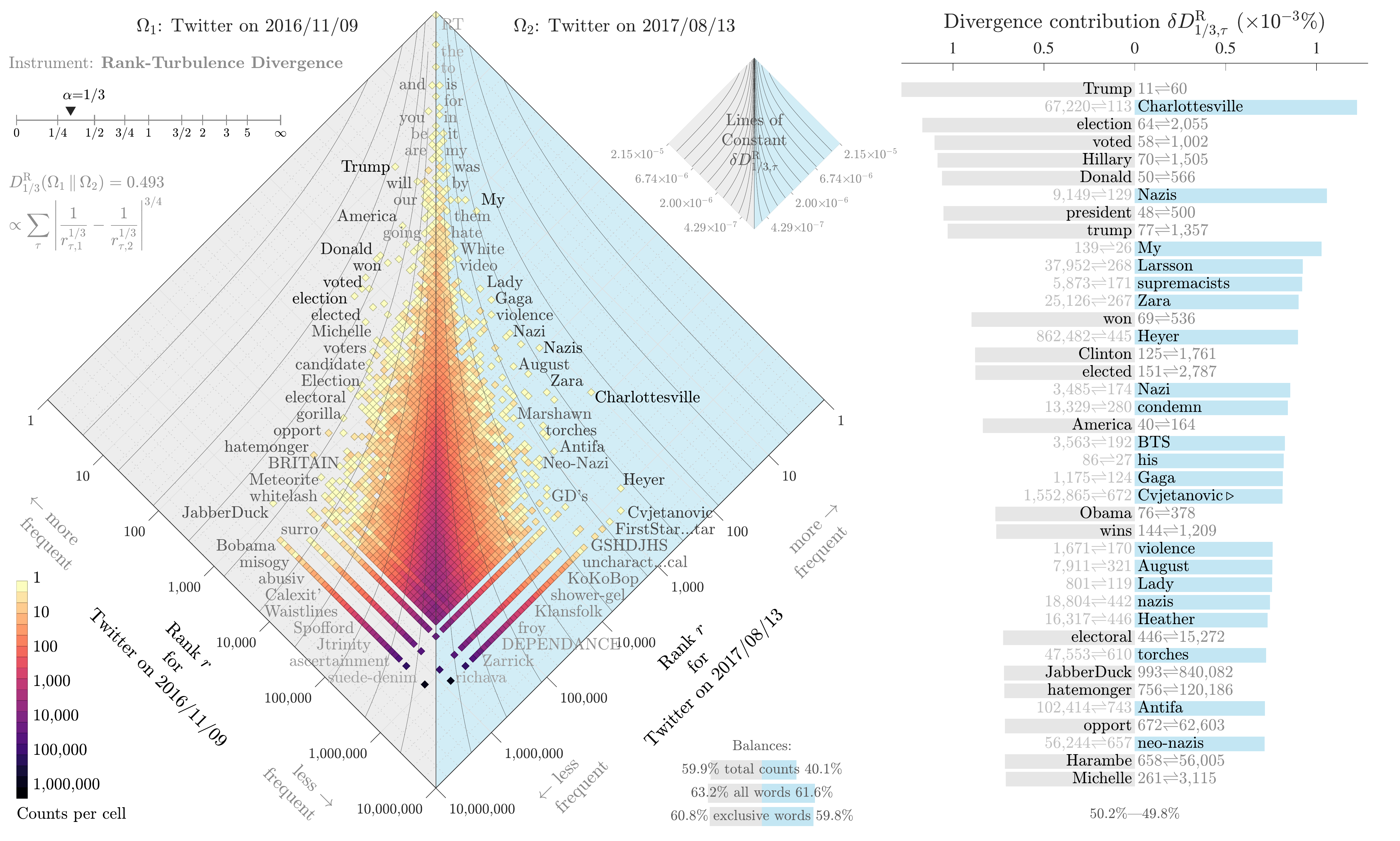}
  \caption{
    \textbf{Example allotaxonograph
    using rank-turbulence divergence to compare word usage on different days of Twitter.}
    We examine the same dates of
    the 2016 US Presidential Election and the Charlottesville Unite the Right rally
    as the rank-rank histogram of
    Fig.~\ref{fig:rankturbdiv.zipfturbulence-2016-11-09-2017-08-13-story-wrangler-all-rank-bare}.
    We add rank-turbulence divergence to
    Fig.~\ref{fig:rankturbdiv.zipfturbulence-2016-11-09-2017-08-13-story-wrangler-all-rank-bare}'s
    histogram with an overlay of contour lines,
    a gauge for $\alpha$ and the expression for $\rtd{1/3}$
    in the upper left corner,
    and a scale for the contour lines in the upper right.
    Based on contributions of each word to $\rtd{1/3}$,
    we generate the ordered list on the right
    by descending values of $\rtdelement{1/3}$.
    Words are arranged left and right and colored gray and blue
    in accordance with the date
    on which they are most prevalent.
    The two dates' ranks for each word in the list are indicated on the opposite side.
    For example,
    $\zipfranktypesystema{Trump}$=11
    and 
    $\zipfranktypesystemb{Trump}$=60,
    and
    $\zipfranktypesystema{Heyer}$=862,482
    and 
    $\zipfranktypesystemb{Heyer}$=445.
    While an exact match is intended, a few annotated words on the histogram differ
    from Fig.~\ref{fig:rankturbdiv.zipfturbulence-2016-11-09-2017-08-13-story-wrangler-all-rank-bare}
    due to chance (e.g., `HURRICANE' and `BRITAIN' on the left side).
    The instrument's function and layout are highly configurable
    in our figure-building script.
    For example,
    the choice of divergence (rank or otherwise),
    axis limits,
    maximum length of type names,
    histogram cell size,
    and
    the guide adornments `less talked about' and `more talked about'
    are all system-specific settings.
    As a design choice, we limit the resolution of $\alpha$ to multiples of 1/12,
    For further details on the underlying histogram,
    see the caption of
    Fig.~\ref{fig:rankturbdiv.zipfturbulence-2016-11-09-2017-08-13-story-wrangler-all-rank-bare}.
  }
  \label{fig:rankturbdiv.allotaxonometer9000-2016-11-09-2017-08-13-story-wrangler-all-rank-div_05}
\end{figure*}

We now combine rank-rank histograms with rank-turbulence divergence
to generate a tunable single-parameter instrument for exploring how two systems differ.
In Fig.~\ref{fig:rankturbdiv.allotaxonometer9000-2016-11-09-2017-08-13-story-wrangler-all-rank-div_05},
we present a `rank-turbulence divergence graph' as an example allotaxonograph.
We again compare the two days of Twitter---the 2016 US election with the 2017 Charlottesville riots---that
we examined in Sec.\ref{subsec:rankturbdiv.viz}.

There are two main components to our general divergence graphs: A map-like histogram
and an ordered list of types contributing the most to the divergence measure being employed.

First we build upon the histogram of
Fig.~\ref{fig:rankturbdiv.zipfturbulence-2016-11-09-2017-08-13-story-wrangler-all-rank-bare}.
We use rank-turbulence divergence with $\alpha = 1/3$, as indicated on the scale in
the top left of the graph.
We discuss the choice of $\alpha$ below.
In all our divergence graphs, we include the divergence's expression above
the top left of the histogram. 
We overlay the histogram with contour lines of constant $\rtdelement{1/3}$.
The contour lines are chosen so that they are evenly spaced and
anchored along the bottom two axes,
making for simple tracking as $\alpha$ is varied.
The inset to the upper right of the histogram provides a scale
for values of $\rtdelement{1/3}$.

For our own implementation of rank-turbulence divergence, we have chosen
to make the increments of $\alpha$ discrete as multiples of 1/12.
This discretization is particularly useful for $\alpha \le 3/2$,
the range of $\alpha$ for which most of the variation in rank-turbulence divergence takes place.
The $\alpha$ scale in the top left 
Fig.~\ref{fig:rankturbdiv.allotaxonometer9000-2016-11-09-2017-08-13-story-wrangler-all-rank-div_05}
shows an inverse tangent transformation that is effective for functional use of the instrument.
As we will see, near $\alpha$=0,
the list's variation with steps of 1/12 is not abrupt.

As they are independent of divergence measures,
the annotations and their locations on the histogram remain unchanged from
Fig.~\ref{fig:rankturbdiv.zipfturbulence-2016-11-09-2017-08-13-story-wrangler-all-rank-bare}.
We now incorporate a linear gray scale based on $\rtdelement{1/3}$,
with higher scoring words accentuated, lower scoring words faded.
We now see `Trump' and `Charlottesville' stand out.
Common words that have not changed rank (`RT', `the', and `to')
as well as words rare on one day and absent on the other
(`suede-denim' and `richava') have all been strongly backgrounded.

Second, we locate a list of words on the right of the instrument in
Fig.~\ref{fig:rankturbdiv.allotaxonometer9000-2016-11-09-2017-08-13-story-wrangler-all-rank-div_05}.
We order the top 40 words by decreasing value of $\rtdelement{1/3}$, indicated by the
underlying bars.
We orient words to the left and right in accordance with the day
of their higher rank;
the bar colors of light gray and light blue match the histogram's format.
Opposite each bar, we show the word's rank on each day.

For example, we see `Trump' has the highest divergence contribution overall,
moving from $\zipfrank$=11 to 60. These ranks indicate a maintenance of
extraordinary levels of lexical ultrafame~\cite{dodds2019a}), but the drop
from $\zipfrank$=11 to 60 registers more strongly for $\rtdelement{1/3}$
than all other rank shifts.
On the opposing date, `Charlottesville' scores comparably to `Trump'
and is second overall.
In contrast to `Trump', however,
`Charlottesville' is a word that changes rank dramatically across the two dates,
moving from $\zipfrank$=67,220 to 113.

For systems for which we are confident we have determined the constituent elements,
it is useful to be able to see which important (i.e., high $\rtdelement{\alpha}$)
elements are part of only one system.
In the ordered list, we indicate types that appear in only of the two systems by a directed open triangle,
that will
either precede a word appearing on the left or trail a word appearing on the right.
For Fig.~\ref{fig:rankturbdiv.allotaxonometer9000-2016-11-09-2017-08-13-story-wrangler-all-rank-div_05}
with $\alpha$ set at $1/3$,
there is only one such word in the top 40 divergence contributions: `Cvjetanovic'.
For general systems,
as we tune $\alpha$ towards zero, more single-system types will move up the list,
and conversely fall back down if we instead dial $\alpha$ towards $\infty$.

In all allotaxonographs, we show three kinds of balances at the bottom right
of the rank-rank histogram.
First, we see the breakdown of total counts between the two dates at 59.9\% and 40.1\%
(the election generated more tweets than Charlottesville).
Second, we have that all words in the lexicon for the two days combined, just over 60\% appear
on each of the two days.
Third, we create separate lexicons for each day,
and find that around 60\% are exclusive for both days, giving a sense of strong turnover.
As we will see, these balances can vary greatly across system comparisons.

\subsection{Tuning Rank-Turbulence Divergence Allotaxonographs}
\label{subsec:rankturbdiv.tuning}

For
Fig.~\ref{fig:rankturbdiv.allotaxonometer9000-2016-11-09-2017-08-13-story-wrangler-all-rank-div_05},
we have chosen $\alpha=1/3$ because it delivers a reasonably balanced list of words with ranks from
across the common-to-rare spectrum.
Our choice here is based purely on a visual inspection.
We have considered several automated methods for determining an optimal $\alpha$,
but leave these for future work.

To demonstrate how tuning $\alpha$ controls the contour lines and alters
the word list
on a rank-turbulence divergence graph, we provide
Flipbook~\flipbooktwitter\
where we sweep through a set of 11 $\alpha$ values in steps:
$0$,
$\frac{1}{12}$,
$\frac{2}{12}$,
$\frac{3}{12}$,
$\frac{4}{12}$,
$\frac{5}{12}$,
$\frac{6}{12}$,
$\frac{8}{12}$,
$1$,
$2$,
$5$,
and
$\infty$.
As we increase $\alpha$,
the set of words (and in general, types) with highest $\rtdelement{\alpha}$ transform
from being dominated by rare words to function words.
Even so, a few words maintain prominence across a wide range of $\alpha$.
For example, `Trump' is the top word for $\alpha$=1/3 to 5/4,
dropping only to 5th for $\alpha$=$\infty$.
(Because of its function-word-like fame,
for $\alpha \le 1/6$, `Trump' does not register in the top 40.)
For $0 \le \alpha \le 5/6$, Charlottesville-related words lead
the right side of the list
(`Cvjetanovic', `Heyer', and `Charlottesville').
At the limit of $\alpha$=$\infty$, the only top 40 Charlottesville word
is `white' (per the prevalence of `white supremacists' and similar terms).

To further our investigation, We provide two more Flipbooks for Twitter.
Flipbook~\flipbooktwitterRT\ shows how the allotaxonograph
of Fig.~\ref{fig:rankturbdiv.allotaxonometer9000-2016-11-09-2017-08-13-story-wrangler-all-rank-div_05}
changes if we control the percentage of retweets included in our sample.
In varying from 1\% to 100\%, we see that the texture of the election side
does not change greatly---the amplified and unamplified versions of Twitter match well.
However, the Charlottesville date shows that the 1\% retweet sample is much more
pop culture focused. As we move through
Flipbook~\flipbooktwitterRT\
and dial up to fully include all retweets for 2017/08/13,
we see words surrounding the events in Charlottesville rise up the list
of dominant contributions.

In Flipbook~\flipbooktwittertimediff, we start with 2019/01/03 and
compare forwards in time, roughly doubling the number of days for each step,
ending with 2020/01/04, the date of the assassination of the Iranian general
Soleimani by the United States.
We see the topics of anchor date 2019/01/03 become more clear as the date moves further into
the past: Government shutdown, the border wall,
and Congresswoman Rashida Tlaib.
The comparison future date travels though a wide range of events.
We observe
that rank-turbulence divergence slowly increases as we compare days increasingly further apart.
Visually, we see the rank-rank histogram broaden subtly.
Determining how an optimal $\alpha$ changes with time scales
would be a natural part of possible future work.

To explore in more depth the value of having a tunable allotaxonometric instrument,
we move away from news and Twitter to consider distributions
presented by two different kinds of systems,
one ecological, the other cultural:
Tree species abundances and popularity of baby names.

\subsection{Species abundance: Example Rank-Turbulence Divergence Allotaxonograph for the limit of $\alpha$=$0$}
\label{subsec:rankturbdiv.alpha0}

\begin{figure*}
  \includegraphics[width=1.1\textwidth,center]{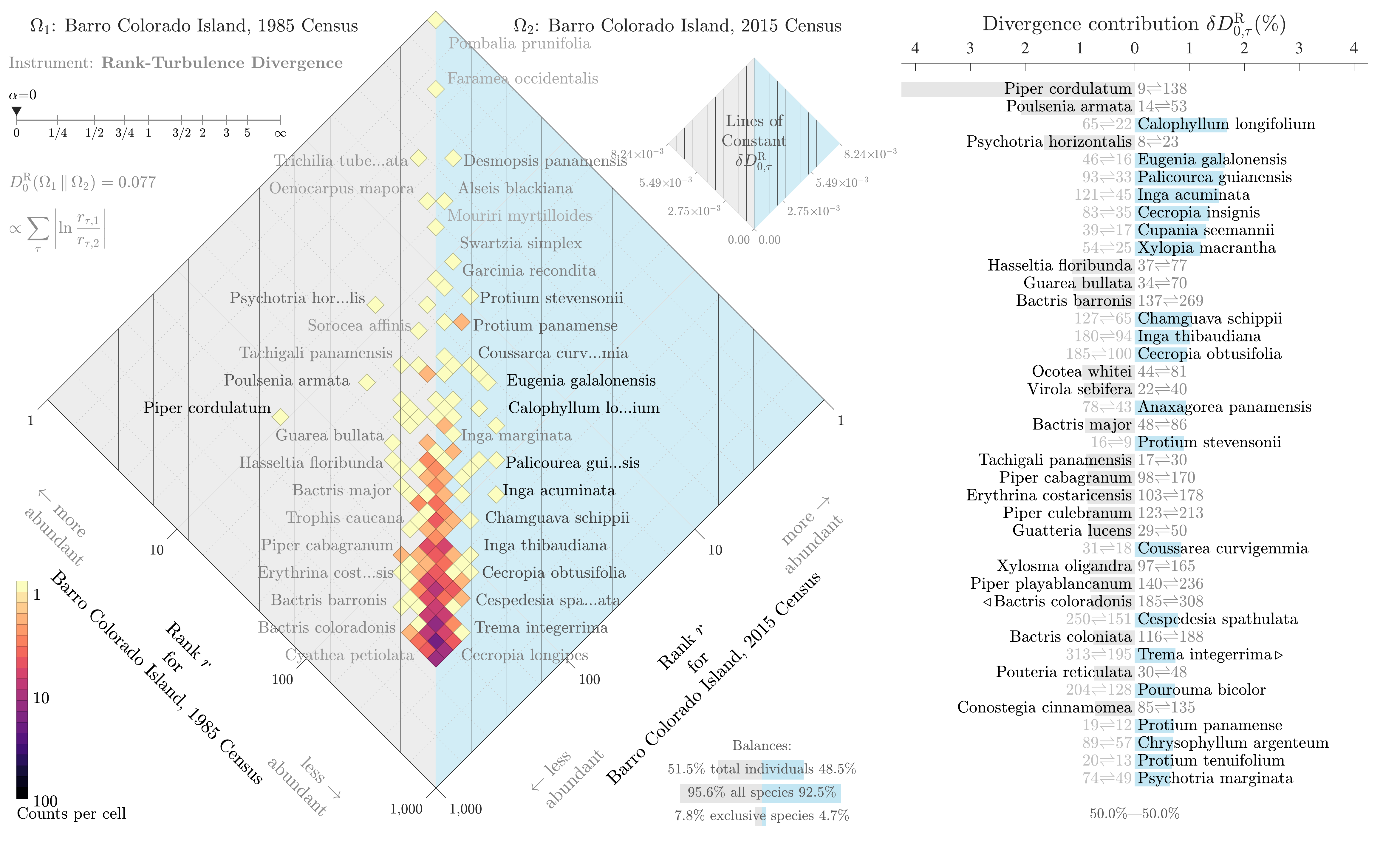}
  \caption{
    \textbf{Allotaxonograph using rank-turbulence divergence
    to compare tropical forest tree species abundance on Panama's Barro Colorado Island (BCI)
    for 5 year censuses completed in 1985 and 2015~\cite{condit2019a}.}
    This system comparison shows relatively little turnover or turbulence.
    We see none of the sideways flaring of the histogram
    towards the bottom---turbulence---as we did for Twitter word usage in
    Fig.~\ref{fig:rankturbdiv.allotaxonometer9000-2016-11-09-2017-08-13-story-wrangler-all-rank-div_05}.
    A choice of $\alpha=0$ for rank-turbulence divergence
    per \Req{eq:rankturbdiv.elementrankturbdiv_zerolimit}
    produces vertical contour lines that conform well to the histogram.
    From inspection of both the histogram
    and
    the $\rtdelement{0}$ list,
    the relative decline of a single species of pepper plant,
    \textit{Piper cordulatum}~\cite{trelease1927a,standley1927a,thies2004a,andrade2013a},
    is the dominant dynamical change in the forest's composition.
    See Sec.~\ref{sec:rankturbdiv.methods.datasets} for further notes on the BCI data.
  }
  \label{fig:rankturbdiv.allotaxonometer9000-1985-2015-barro-colorado002}
\end{figure*}

In Fig.~\ref{fig:rankturbdiv.allotaxonometer9000-1985-2015-barro-colorado002},
we show a rank-turbulence divergence graph comparing tropical tree species numbers
on Barro Colorado Island (BCI) in the Panama Canal~\cite{condit2000a} for five-year censuses
completed in
1985 
and
2015 ($\systema$ and $\systemb$)~\cite{condit2019a}.

In being visually close to the limit of comparing two identical rankings
(Fig.~\ref{fig:rankturbdiv.zipfturbulence-2016-11-09-2017-08-13-story-wrangler-all-rank-bare}B),
the histogram's vertical linear form
immediately shows
that the species abundance distributions are strongly aligned.
Because of the possibility of exogenous catastrophic events such as fires and
the abrupt transitions accessible by complex dynamical systems~\cite{strogatz1994a},
the composition of an ecological system may change dramatically over a few decades.
For this example from BCI, however, we see a system that is strongly durable in its
component rankings.

We compare the 1985 and 2015 distributions by applying
rank-turbulence divergence with $\alpha = 0$.
The overall score
$\rtdalphavarsystemsRank{0} = 0.077$
is well short of the randomized equivalent
of
$\rtdalphavarsystemsRankRand{0} = 0.376$
(from 100 samples; standard deviation $\sigma$=0.012).
Per
\Req{eq:rankturbdiv.elementrankturbdiv_zerolimit},
the contribution to overall divergence by
changes in species abundance follows a
log-ratio of ranks:
$
\left\lvert
\ln
\zipfrank_{\elementsymbol,\indexaraw}/\zipfrank_{\elementsymbol,\indexbraw}
\right\rvert.
$
The contour lines for
constant $\rtdelement{0}$ accord with the histogram's form.
From the histogram and $\rtdelement{0}$ list,
we see one species of pepper plant---\textit{Piper cordulatum}~\cite{trelease1927a,standley1927a,thies2004a,andrade2013a}---stands out,
having diminished markedly in relative abundance,
dropping from
$\zipfrank_{\indexaraw}$=9
to
$\zipfrank_{\indexbraw}$=138.
Two other species that have dropped in relative abundance
feature in the top 4 of the $\rtdelement{0}$ list:
\textit{Polsenia armata} 
($\zipfrank_{\indexaraw}$=14
to
$\zipfrank_{\indexbraw}$=53)
and 
\textit{Psychotria horizontalis} 
($\zipfrank_{\indexaraw}$=8
to
$\zipfrank_{\indexbraw}$=23).

Per the balance indicators,
we see that the total number of individuals in each year's census
is roughly the same (51.5\% and 48.5\%),
that most types for both years appear in each
system (95.6\% and 92.5\%),
and that relatively few types are exclusive to each year
(7.8\% and 4.7\%).
Only two year-exclusive species make the top 40 for
$\rtdelement{0}$ contributions:
\textit{Bactris coloradonis} (1985 only)
and
\textit{Trema integerrima} (2015 only).
Regarding changes in overall diversity,
we see that the loss of \textit{Piper cordulatum}
has not been to the gain of a single species---there is no
one species on the right of the histogram with a distinctly high
$\rtdelement{0}$.
Of the top 10 species ranked by $\rtdelement{0}$,
7 are species that have become relatively more abundant.
For the top 40, the balance is 20 down and 20 up.
Overall, our instrument's dashboard makes clear that
there is a singular drop in \textit{Piper cordulatum}'s
ecological role amid incremental (and possibly also important) changes for other species,
straightforwardly directing future research attention.

\subsection{Baby names: Example Rank-Turbulence Divergence Allotaxonograph for the limit of $\alpha$=$\infty$}
\label{subsec:rankturbdiv.alphainfty}

\begin{figure*}
                    \includegraphics[width=1.1\textwidth,center]{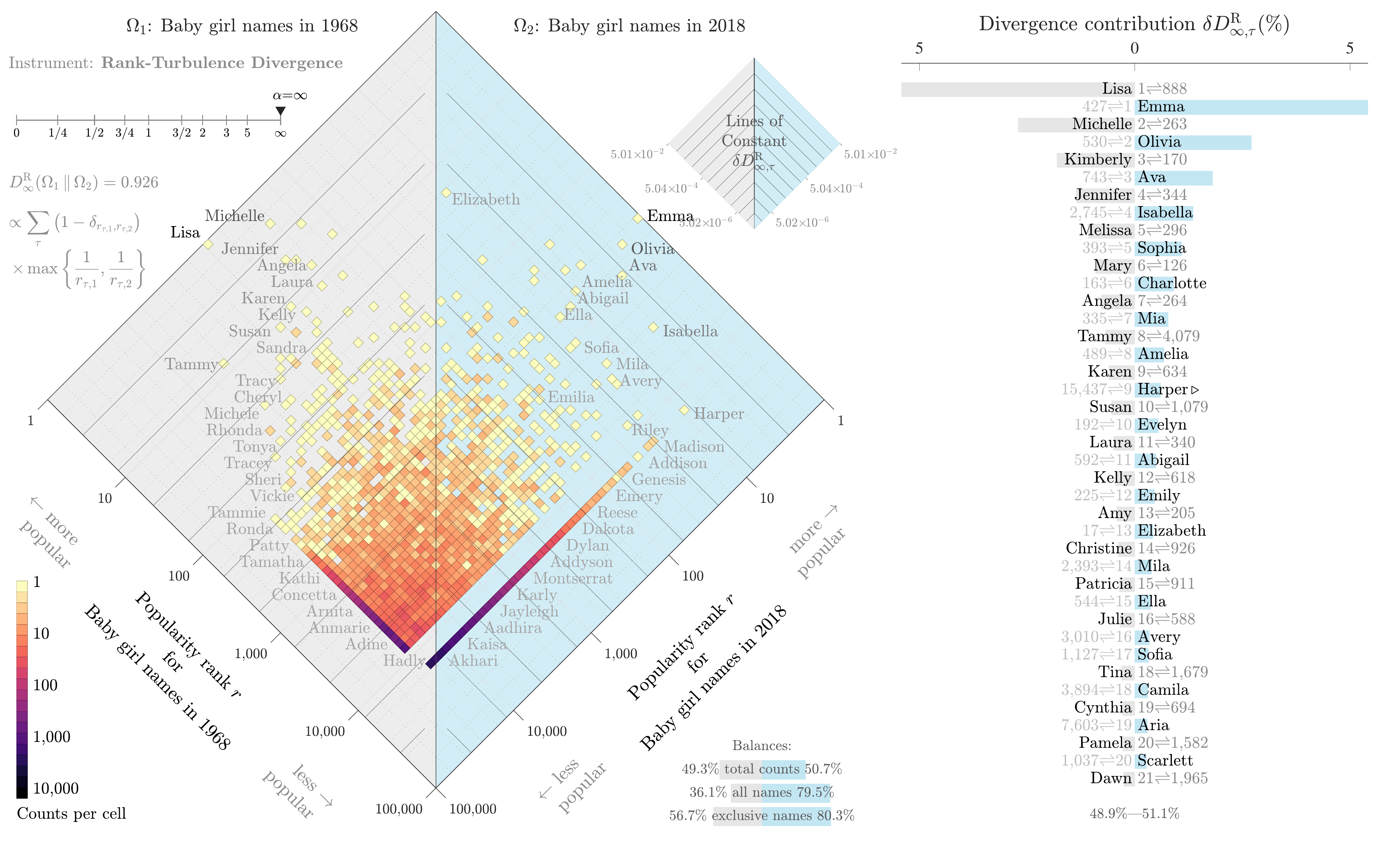}
   \caption{
    \textbf{Allotaxonograph comparing names of girls born in the US in 1968 and 2018.}
    For dataset details, see Sec.~\ref{sec:rankturbdiv.methods.datasets}.
    Of our four main case studies, baby name distributions show the strongest change
    with $\rtdalphasystemsOmega$ scores verging on that of the random equivalent.
    The asymmetry of the separated 2018-exclusive names and the balance score of
    80.3\% of all names in 2018 being new relative to 1968
    show that while there is much social imitation (see 1970s, `Jennifer'),
    baby names are highly innovative collectively.
    Note that at the bottom of the histogram, `Hadly' is a 2018 exclusive word
    but it is oriented towards the left per our annotation method
    (see Fig.~\ref{fig:rankturbdiv.zipfturbulence-2016-11-09-2017-08-13-story-wrangler-all-rank-bare}
    and Sec.~\ref{subsec:rankturbdiv.viz}).
    See Fig.~\ref{fig:rankturbdiv.allotaxonometer9000-babynames-boys} for the boy name version.
    For 1968--2008,
    Flipbook~\flipbookgirlsyears\
    shows how the list of contributions to rank-turbulence divergence changes
    as $\alpha$ varies from 0 to $\infty$.
    Flipbook~\flipbookgirlsalphas\
    provides a sweep of $\alpha$=$\infty$
    allotaxonometric graphs for girl names over time,
    for 50 year gap comparisons starting
    with 1880--1930 and moving forward in 5 year steps.
  }
  \label{fig:rankturbdiv.allotaxonometer9000-babynames-girls}
\end{figure*}

\begin{figure*}
                    \includegraphics[width=1.1\textwidth,center]{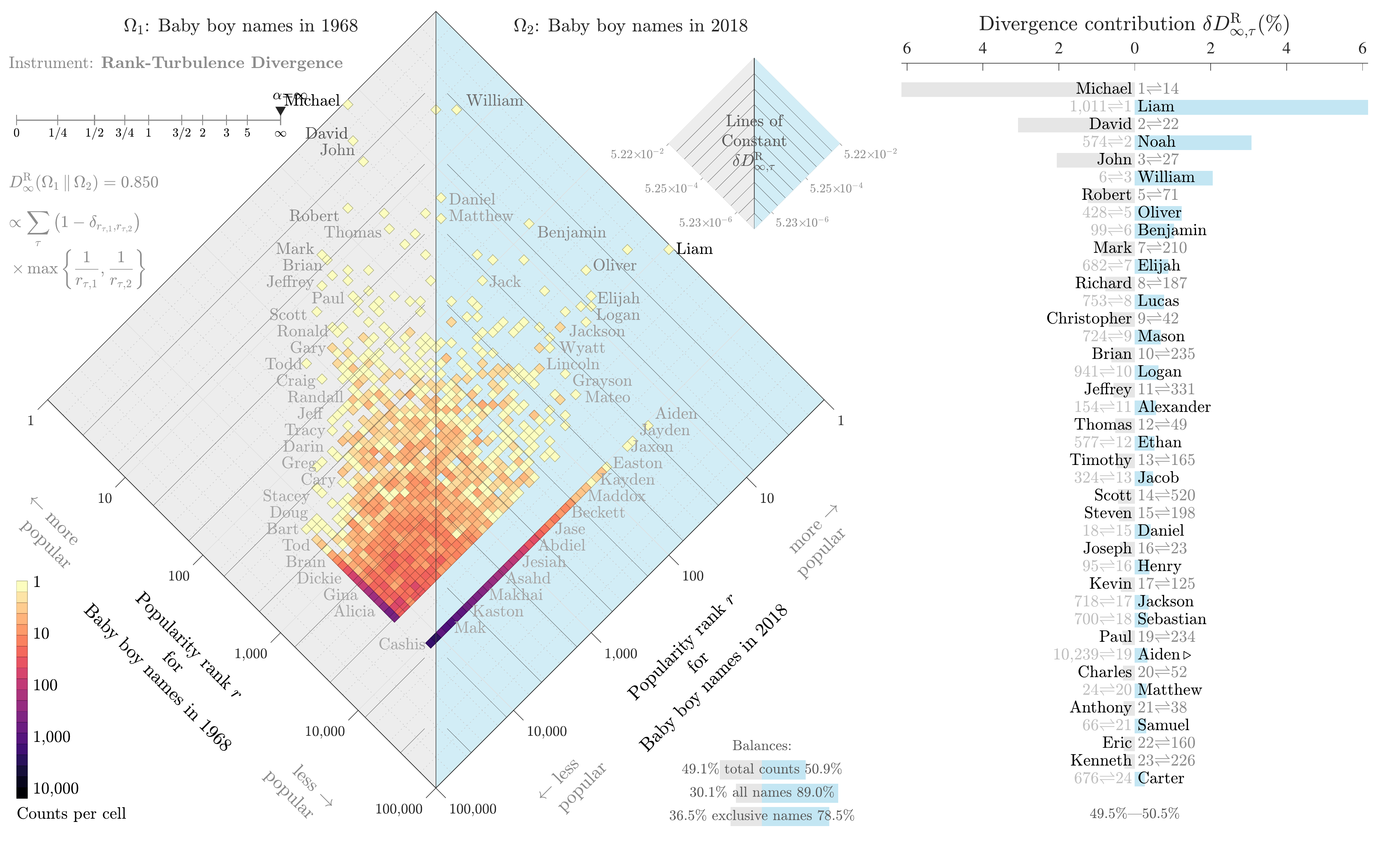}
  \caption{
    \textbf{Allotaxonograph comparing US boy names for the years 1968 and 2018.}
    For dataset details, see Sec.~\ref{sec:rankturbdiv.methods.datasets}.
    At the bottom of the histogram,
    `Cashis' is oriented to the left but is a 2018 exclusive word,
    is per `Hadly' in Fig.~\ref{fig:rankturbdiv.allotaxonometer9000-babynames-girls}.
    As for girl names, we provide two Flipbooks showing
    50 year gap comparisons moving through time (Flipbook~\flipbookboysyears)
    and the 
    effects of varying $\alpha$ for the 1968--2018 comparison
    (Flipbook~\flipbookboysalphas).
  }
  \label{fig:rankturbdiv.allotaxonometer9000-babynames-boys}
\end{figure*}

Now, for an example of where tuning rank-turbulence divergence's
parameter $\alpha$ to the limit of $\infty$ is helpful,
we explore the temporal evolution of US baby name popularity~\cite{hahn2003a,wattenberg2005a}.
Because of the richness of baby name trends, we will also show
how the full range of $\alpha$ can be used to uncover cultural changes.
The dataset we use
tabulates annual name frequencies running from 1880 through to 2018,
and is derived from Social Security card applications.
For privacy, there is a truncation instituted in the dataset,
and only baby names for which there are 5 or more instances in a year
are included, and our analysis and discussion. 
As for any complex system where quantification is imperfect,
our discussion and analysis below carries the caveat that
apparent system exclusive types may in fact be present
(for further details and limitations see
Sec.~\ref{sec:rankturbdiv.methods.datasets}).

In Fig.~\ref{fig:rankturbdiv.allotaxonometer9000-babynames-girls},
we use a
rank-turbulence divergence graph
with $\alpha$=$\infty$
to compare changes in baby name frequencies for girls born in the US in 1968
and girls born in the US in 2018, a 50 year gap.
In Fig.~\ref{fig:rankturbdiv.allotaxonometer9000-babynames-boys},
we present the corresponding allotaxonomic graph for boy names.
In the \suppmaterial, we provide
 Flipbooks with $\alpha$=$\infty$
showing half century changes for both
girl and boy names starting in 1880 and moving forward in 5 year increments
(Flipbooks~\flipbookgirlsyears\
and
\flipbookboysyears),
as well as
Flipbooks for the same 1968--2018 comparison with $\alpha$ varying from 0 to $\infty$
(Flipbooks~\flipbookgirlsalphas\
and
\flipbookboysalphas).
For baby names, an interactive version of the instrument would allow
tunable $\alpha$ and the choice of years to be readily explorable.

In contrast to the lexical turbulence of Twitter
and the largely vertical form we saw for forest species counts,
the histograms in
Figs.~\ref{fig:rankturbdiv.allotaxonometer9000-babynames-girls}
and~\ref{fig:rankturbdiv.allotaxonometer9000-babynames-boys}
bear strong signatures of randomness and innovation.

First, as we saw in
Fig.~\ref{fig:rankturbdiv.zipfturbulence-2016-11-09-2017-08-13-story-wrangler-all-rank-bare}C,
a random shuffling of ranked lists results in histograms
predominantly weighted in the lower triangle of the plot.
We see a strong imprint of this limiting case in
Figs.~\ref{fig:rankturbdiv.allotaxonometer9000-babynames-girls}
and~\ref{fig:rankturbdiv.allotaxonometer9000-babynames-boys},
reflective of a great deal of cultural and societal change.

Second, we see dense exclusive-type lines at the base of both sides
of the histograms 
in Figs.~\ref{fig:rankturbdiv.allotaxonometer9000-babynames-girls}
and~\ref{fig:rankturbdiv.allotaxonometer9000-babynames-boys},
the stamp of disjoint systems
(Fig.~\ref{fig:rankturbdiv.zipfturbulence-2016-11-09-2017-08-13-story-wrangler-all-rank-bare}D).
The asymmetry of the histograms, with the separated exclusive-type line
on the lower right,
reflects
the strong innovation of 2018 names relative to 1968.
Overall, the turnover is stronger for girl names than boy names.
We can get a sense of this visually by observing
that there is less flare to the left of the histogram for boy names
relative to the histogram for girl names.
The balance quantities show one major difference:
For girls, 56.7\% of 1968 names are exclusive to 1968
while for boys, the same quantity is only 36.5\%.

For girls, ranging from common 2018 names
(`Harper',
`Madison',
and
`Addison')
down to rare names
(`Kaisa',
`Akhari',
and
`Hadly'),
the 2018 exclusive names comprise 80.3\% of all names
(14,485 of 18,029).
For the smaller name base of boys,
we see 10,994 of 14,004 (78.5\%) names are 2018 exclusive.
Not registering above 5 counts in 1968 but widespread
in 2018 are
`Aiden',
`Jaxon',
and
`Maddox',
and three 2018 exclusive but rare examples are
`Kaston',
`Mak',
and
`Cashis'.

While not separated because of the histogram's cell sizes,
the 1968 exclusive-type line is dense relative to the histogram body
in both 
Figs.~\ref{fig:rankturbdiv.allotaxonometer9000-babynames-girls}
and~\ref{fig:rankturbdiv.allotaxonometer9000-babynames-boys}.
We find
56.7\% of all girl names (4,650 of 8,194)
and
36.5\% of all boy names (1,732 of 4,742)
are 1968-exclusive names relative to 2018.
A wide range of girl names that were popular in 1968
(`Tammie', `Ronda', and `Patty')
as well as rare
(`Anmarie' and `Adine')
have fallen out of favor by 2018.
For boys, once-common `Bart' and `Tod' have dropped off the ledger.
We also see apparent errors along the
exclusive-type line for boy names in 1968
with `Gina' (20 counts) and `Alicia' (9 counts).

We note that the asymmetries of both histograms---their apparent
right-side `heaviness'---are not due even in part
to changes in overall numbers.
The total number of girl names recorded in 1968 and 2018
are comparable at 1,709,551 and 1,846,101 (7.99\% increase);
for boys, these numbers are 1,775,997 and 1,928,871 (8.61\% increase).
The number of unique names in the years 1968 and 2018
are strikingly different however:
8,194 and 18,029 for girls (120\% increase),
and 4,742 and 14,004 for boys (195\% increase).
Two of the major factors which lead to this explosion in name-space
are immigration and creation of new names.

Using the overall birth numbers, we can estimate the percentage
of names absent from our dataset---those with less than 5 instances:
4.06\% for 1968 and 8.62\% for 2018 for girls,
and 
2.11\% for 1968 and 6.66\% for 2018 for boys.
The 2018 Zipf distributions thus have heavier tails
pointing once again to strong innovation.

The turnover in girl names results in a high
rank-turbulence divergence value of 
$\rtdalphavarsystemsRank{\infty} = 0.926$.
For the same time frame comparison,
boy names have a lesser but still high value of
$\rtdalphavarsystemsRank{\infty} = 0.850$.
Both values are below but not far from
the randomized equivalents 
with Zipf distributions held constant:
$\rtdalphavarsystemsRankRand{\infty} = 0.973$
and 0.966.

We turn to the overall orderings of $\rtdelement{\infty}$
contributions for girls and boys,
the ordered lists of
Figs.~\ref{fig:rankturbdiv.allotaxonometer9000-babynames-girls}
and~\ref{fig:rankturbdiv.allotaxonometer9000-babynames-boys}.

In general, in the limit of $\alpha$=$\infty$, the contribution ordering
will be an interleaving of types from both distributions.  The
ordering of types on each side of the list will match those of the
separate Zipf distributions with the exception that all types that do
not change rank will be absent.  The interleaving is generally a
simple back and forth sequence between the two systems but breaks
whenever a rank is reached that is the maximum rank for a specific
type.

For girls in 1968 relative to 2018,
we see the three medal places go to
`Lisa',
`Michelle',
and `Kimberly'.
In fourth, we have `Jennifer', 
a name that would go on to be the most popular girl name in the US throughout the entire 1970s.
In fifth is the once dominant `Mary' which had held the number one position from 1880 through to 1961.
 
The dominance of the most popular girl name in 1968, `Lisa', relative to
2018 is remarkable, carrying the top overall 1968 $\rtdelement{\infty}$ contribution
for all values of $\alpha$.
In Flipbook~\flipbookgirlsalphas,
we see that in dropping from
$\zipfrank$=1
to
$\zipfrank$=888,
`Lisa' is second in contribution for both 1968 and 2018 
only for $\alpha=0$ (first page)
when we see `Harper' take the top position.
At this limit, order is by rank ratio
and the above-the-rim elevation
for `Harper' from 
$\zipfrank$=15,437
to
$\zipfrank$=9
is more than enough for the win.

On the other side, for 2018 relative to 1968, `Emma' is the new `Lisa',
with `Olivia' and `Ava' in second and third
for $\rtdelement{\infty}$ contribution.
In dialing $\alpha$, Flipbook~\flipbookgirlsalphas\
shows that like `Lisa',
`Emma' prevails above all other names except
`Harper' when $\alpha=0$.

For boy names, the 1968 $\rtdelement{\infty}$ side of the list
is headed by 
`Michael',
`David',
`John',
and
`Robert'
while for 2018, the top differential names are
`Liam',
`Noah',
`William',
and
`Oliver'.
As we tune $\alpha$ down from $\infty$ to 0
(Flipbook~\flipbookboysalphas),
we see that `Liam' has the top
$\rtdelement{\infty}$ contribution across all $\alpha$,
exceeding the ranges of `Lisa' and `Emma'.

Of special note is the name `Elizabeth' which stands out on the rank-rank histogram, well
isolated in the upper triangle.  We see that of all the top girl names
in 1968, `Elizabeth' alone has held its popularity.
Flipbook~\flipbookgirlsyears,
further shows that `Elizabeth' maintains this isolated stability over decades.
No standard divergence measure will highlight `Elizabeth', inviting the
development of a different class of measures that find anomalous
rank-rank pairs.

While not to the degree of `Elizabeth',
there are two boy names that occupy
a small hollowed-out region of rank-rank space in the
histogram
of Fig.~\ref{fig:rankturbdiv.allotaxonometer9000-babynames-boys}:
`James' (steady at $\zipfrank$=4)
and
`William'
(up from $\zipfrank$=6 to $\zipfrank$=3).
As `Liam' is an Irish variant on `William',
the latter effectively held the 1st and 3rd position in 2018.

For girl names compared with the $\alpha$ set to 0,
the first page of Flipbook~\flipbookgirlsyears\
shows that 1968 and 2018 exclusive names dominate the overall list.
While `Lisa' remains at the top, we then
have `Tammy',
`Michele',
`Rhonda',
`Michelle'
and
`Tammie'
as the 6 names from 1968 in the top 40 for $\rtdelement{0}$ contributions.
After `Harper',
the top 2018 names are
`Madison',
`Isabella',
`Luna',
and
`Layla'.

Using $\alpha=0$ for boy names, 
we see in Flipbook~\flipbookboysalphas,
that only one name from 1968 make the top 40 for $\rtdelement{0}$ contributions:
`Bart'.
The top 40 list is otherwise all boy names from 2018,
leading with
`Liam',
`Aiden',
`Jayden',
`Noah',
and
`Jaxon'.

Finally, our allotaxonomic instrument has the ability to uncover
subsets of related types behaving in similar ways.
For example, when tuning to $\alpha$=0
(Flipbook~\flipbookgirlsyears),
we see a raft of 2018 exclusive boy names ending
in `-aden', `-aiden', and `-ayden'.
Investigating further, we find 175 names
appearing 5 or more times in 2018
that are exclusive to 2018 relative to 1968
and matching the regular expression /[Aa][iy]*d+[aeoiuy]n+\$/.
A selection of examples ranging from common to rare,
 highlighting variations on Brayden, are:
`Aiden'	($\zipfrank$=19)
`Jayden'	(30),
`Brayden'	(84),
`Kayden'	(97),
`Zayden'	(185.5),
`Rayden'	(683),
`Braydon'	(856),
`Braiden'	(1,239),
`Bradyn'	(1,936),
`Grayden'	(1,936),
`Braydan'	(3,534.5),
`Braydin'	(3,817.5),
`Bladen'	(4,974.5),
`Blayden'	(5,177),
`Braidyn'	(5,177),
`Vayden'	(5,870),
`Braydyn'	(6,873),
`Wayden'	(7,322),
`Bradon'	(8,434.5),
`Slayden'	(8,434.5),
`Xzayden'	(10,155.5),
 Blaiden'	(11,389.5),
`Braydenn'	(13,042),
and
`Braidon'	(13042).

For girl names, using a similar analysis
for the ending -lyn,
we find 535 names exclusive to 2018,
the top four of which are:
`Adalynn'	($\zipfrank$=108),
`Adalyn'	(144),
`Adelyn'	(226),
and
`Adelynn'	(316)
(there are 21 other names matching
the pattern /\^{}A[aeiouy]*d+[aeiouy]l+[yi]+n+\$/).
There are 85 names exclusive to 1968
that are of the -lyn family
led by
`Jerilyn' ($\zipfrank$=1,152.5),
`Jacalyn' (1,528.5),
and
`Cherilyn' (1,870.5),
and 75 that appear in both 1968 and 2018
(e.g., `Carolyn' and `Evelyn').

These small interrogations of the data lead to larger questions
which are beyond the scope of our work here.
Are girl and boy names differently diverse?
And how has the phonetic spread of names changed over time?
A complete analysis could be performed by matching and grouping names based
on spelling and syllables.

\subsection{Allotaxonometry of publicly traded US companies: Stability, shocks, and errors}
\label{subsec:rankturbdiv.shocks}

\begin{figure*}
  \includegraphics[width=1.1\textwidth,center]{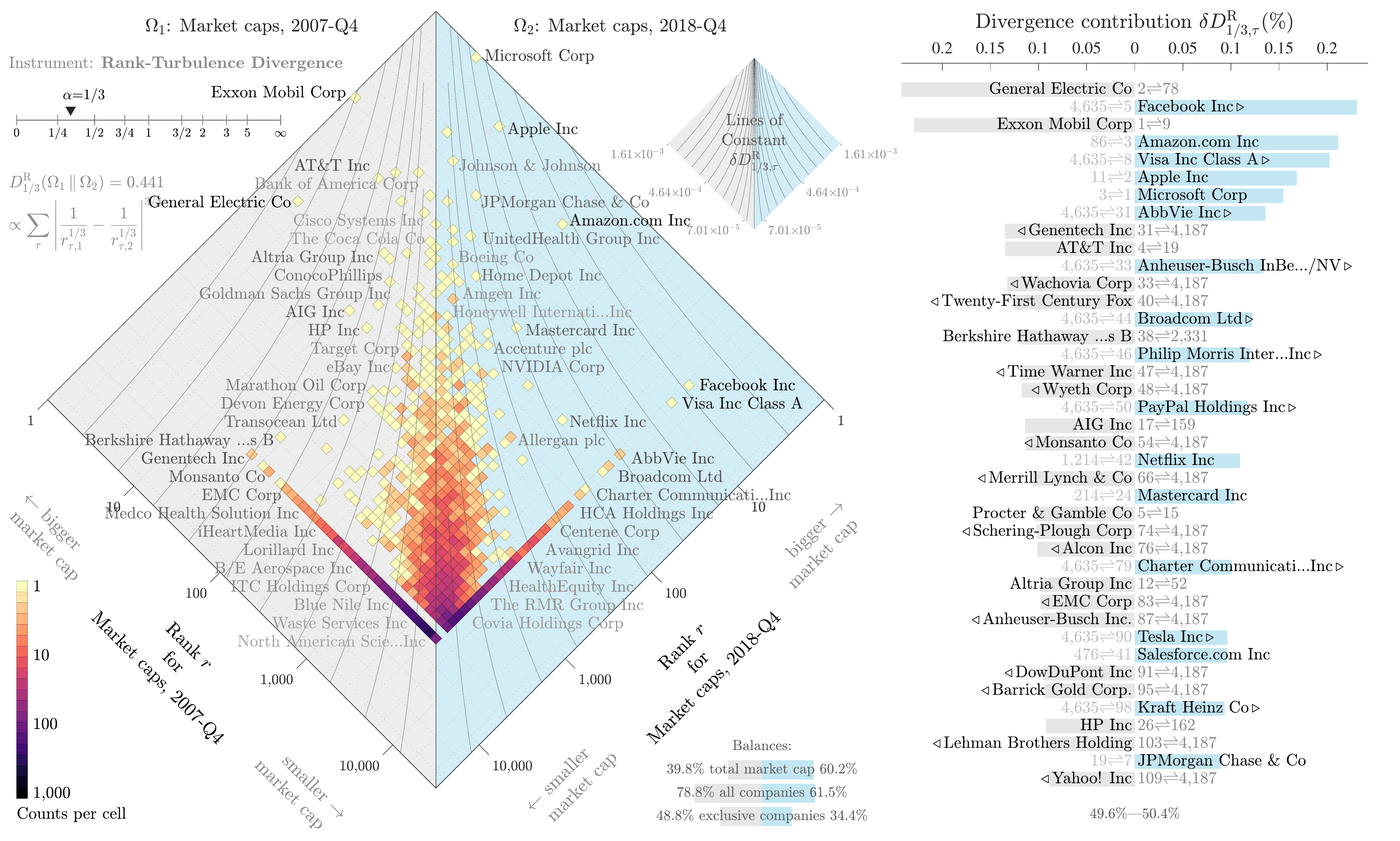}
  \caption{
    \textbf{Allotaxonometric comparison of publicly traded US companies
    in 2007 and 2018 by fourth quarter market capitalization.}
    The rank-rank histogram is a hybrid of a
    vertical structure we see for relatively stable systems 
    (Fig.~\ref{fig:rankturbdiv.zipfturbulence-2016-11-09-2017-08-13-story-wrangler-all-rank-bare}B),
    and a `vee' of disjoint systems 
    (Fig.~\ref{fig:rankturbdiv.zipfturbulence-2016-11-09-2017-08-13-story-wrangler-all-rank-bare}D).
    The disjoint feature results from sharp transitions 
    as companies fail, merge with or are acquired by others, or go public or return to private,
    but also from missing and misrecorded data.
    Berkshire Hathaway's market cap, for example, was misrecorded as a thousand fold drop.
    We include Berkshire Hathaway and other errors in part to show how an
    allotaxonometric analysis can sharply reveal dataset problems.
    See Sec.~\ref{subsec:rankturbdiv.truncation} for discussion,
    and  Sec.~\ref{sec:rankturbdiv.methods.datasets} for dataset details.
  }
  \label{fig:rankturbdiv.allotaxonometer9000-siblis_2007-Q4_2018-Q4_marketcaps004}
\end{figure*}

In Fig.~\ref{fig:rankturbdiv.allotaxonometer9000-siblis_2007-Q4_2018-Q4_marketcaps004},
we show the rank-turbulence divergence graph
comparing US company by market caps
in the final quarter of 2007
with the final quarter of 2018
(for dataset description, see Sec.~\ref{sec:rankturbdiv.methods.datasets}).
The allotaxonograph
is a blend of the two limiting cases of stability and change:
the vertical line of matching systems
and the `vee' of disjoint systems
(Figs.~\ref{fig:rankturbdiv.zipfturbulence-2016-11-09-2017-08-13-story-wrangler-all-rank-bare}B
and
\ref{fig:rankturbdiv.zipfturbulence-2016-11-09-2017-08-13-story-wrangler-all-rank-bare}D).
We choose $\alpha=1/3$ for the rank-turbulence divergence instrument
as the ordering of $\rtdelement{1/3}$ values presents
a mixture of high to low market cap (see below for more on this choice).
In Flipbook~\flipbookmarketcapsyears, we show allotaxonographs
for market cap comparisons for 6 year time gaps
starting 1995 and moving through to 2012.

Of the companies which both existed and
reported market cap in both 2007 and 2018,
we see a great deal
of durability to their rankings.
Somewhat more than what
we see for species abundance numbers in Sec.~\ref{subsec:rankturbdiv.alpha0},
there are some
notable movements in ranks.
At the top of the rank-losing side of
$\rtdelement{1/3}$ list we
see
General Electric ($\zipfrank$=2$\rightarrow$78),
Exxon Mobil (1$\rightarrow$9),
and
AT\&T (4$\rightarrow$19).
Berkshire Hathaway's apparent drop
stems from a dataset error which we discuss below.
On the right side for companies in existence in both 2007 and 2018,
technology companies dominate:
Amazon ($\zipfrank$=86$\rightarrow$3),
Apple (11$\rightarrow$2),
Microsoft (3$\rightarrow$1),
and
Netflix (1,214$\rightarrow$42).

Companies along the exclusive lines of the disjoint system `vee' disappear and appear
for a range of reasons.
Mergers and acquisitions, companies being
taken from public to private and vice versa,
and outright failure
all contribute to market cap comparisons
having a disjoint aspect.

Looking through the 2007 exclusive companies on the histogram and the list
(as indicated by the left triangle prefix),
we see many companies that were acquired,
with a few examples being
Wachovia (bought by Wells Fargo in 2008),
Genentech (bought by Roche in 2009),
Time Warner (bought by Charter Communications in 2016),
and
Monsanto (bought by Bayer, 2018).
We also find a few companies that failed with
Lehman Brothers being a famous (or infamous) example
from the 2007-2008 global financial crisis.

On the 2018 side, Visa and Facebook are the standout entrants.
With respective initial public offerings (IPOs) in 2008 and 2012,
we find them rank at $\zipfrank$=5 and 8 at the end of 2018.
Visa's competitor Mastercard was already publicly traded in 2007,
and ranks highly as well for $\alpha=1/3$
($\zipfrank$=1,214$\rightarrow$24).
AbbVie, Abbot Laboratories in 2013 ranks
highest for pharmaceutical companies.
The brewing company Anheuser-Busch InBev SA/NV
formed in 2008 when Belgium's InBev purchased
Anheuser-Busch.

The dataset for market caps does
have some missing and erroneous data.
DowDuPont's market cap for the last quarter of 2018 is absent
and is consequently shown to have plummeted from
a rank of
$\zipfrank$=91 in 2007
to equal-to-last in 2018.
Berkshire Hathaway's
market cap is clearly misrecorded
for the last three
quarters of the dataset
(apparently dropping from
\$528,336.12M
to
\$335.04M at the end of 2018).

We have chosen to leave such errors in
Fig.~\ref{fig:rankturbdiv.allotaxonometer9000-siblis_2007-Q4_2018-Q4_marketcaps004}
to help demonstrate the importance of using a rich,
graphical allotaxonometric instrument.
With a naive measurement of divergence, we would easily miss problematic data points.
Evidently, further cleaning of the market cap
dataset would be required for further investigations.

The market cap histogram shows the importance of
using a rich allotaxonometric instrument,
and how we must take care when measuring divergences of any kind.
The histogram's form is not as simple as those we have seen for Twitter, species abundance, and baby names,
and it would be evidently problematic to allow for
an unexamined, automated fitting of $\alpha$ for rank-turbulence divergence
(or parameters of any other divergence).

Given the composite form of the allotaxonograph
for market caps, an alternative treatment would be to separate
out companies that appear in both systems from those companies
that appear in only one year.
The enduring companies could be analyzed as a 
low-turbulent system on its own,
and the companies exiting and entering as a disjoint system.
A rank-based divergence instrument could be constructed
that achieves this automatically, possibly returning a set of
measurements that would capture that stable-shock balance we so
clearly observe.
Handling mergers, acquisitions, and partitionings of companies
is also plausible and would require other kinds of elaboration of
rank-turbulence divergence.

\subsection{Truncation Effects for Rank-Based Allotaxonographs}
\label{subsec:rankturbdiv.truncation}

\begin{figure*}
  \includegraphics[width=\textwidth,center]{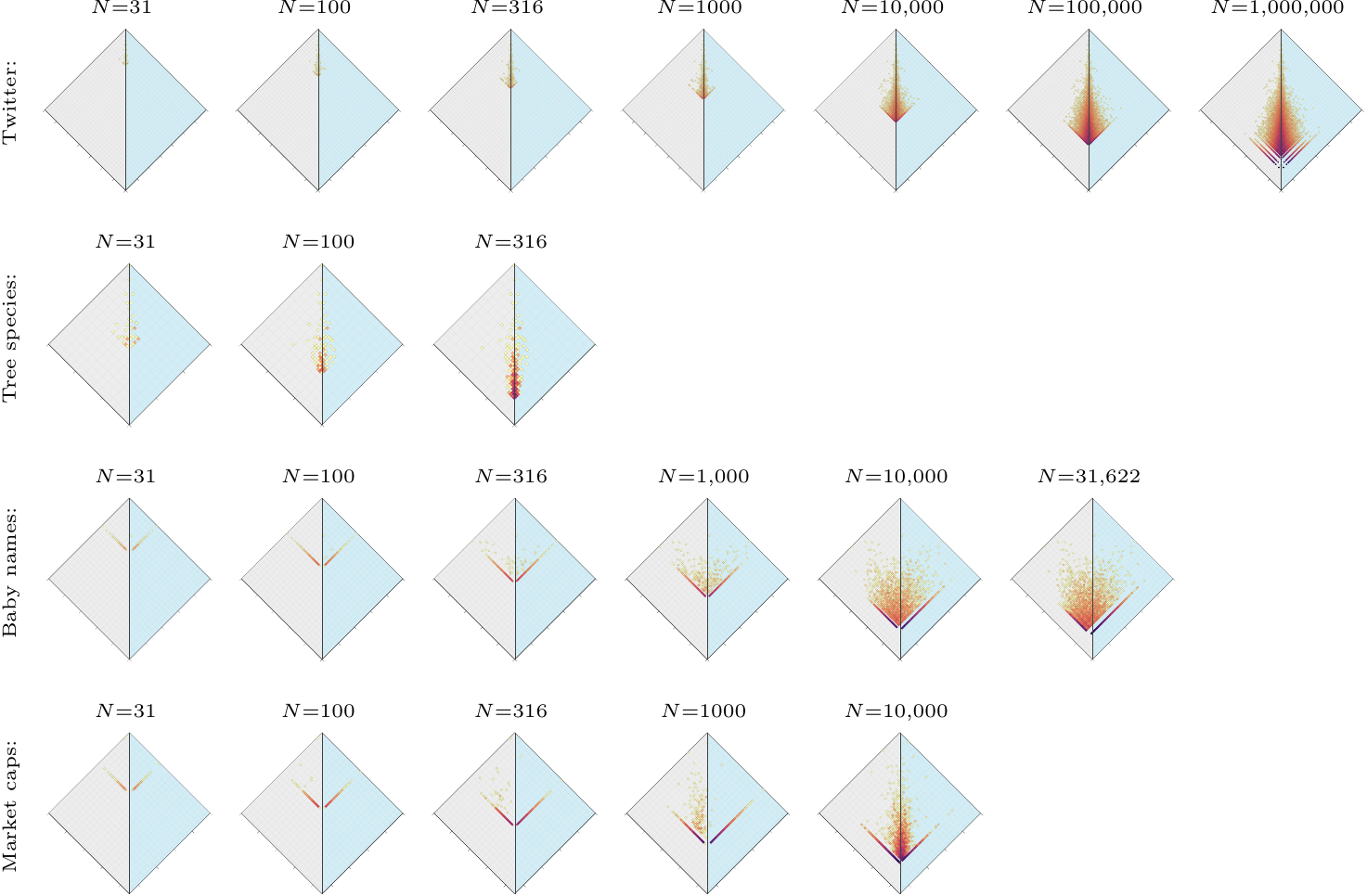}
  \caption{
    \textbf{Exploration of the effect of subsampling data for allotaxonometric analyses.}
    The rows
    correspond to the four case studies of
    Twitter, trees, baby names, and market caps
    (see
    Figs.~\ref{fig:rankturbdiv.allotaxonometer9000-2016-11-09-2017-08-13-story-wrangler-all-rank-div_05}--\ref{fig:rankturbdiv.allotaxonometer9000-siblis_2007-Q4_2018-Q4_marketcaps004}).
    Each row
    shows abstracted rank-rank histograms for Zipf distribution
    truncations to the top $N$ types.
    As $N$ increases, the Twitter and tree species histograms are revealed
    in a clean fashion, while baby names and market caps begin with
    a disjoint system `vee' that masks their large $N$ forms.
    Rows extend to the maximum system size for each comparison,
    and all colormaps and limits correspond to those used for
    the four case studies.
        For allotaxonometric analyses,
    see Flipbooks~\flipbooktwittertrunc---\flipbookcompaniestrunc.
  }
  \label{fig:rankturbdiv.truncation}
\end{figure*}

Truncation of a system's Zipf distributions
is a common if often overlooked problem~\cite{haegeman2013a,koplenig2019a}.
datasets may be curtailed for many reasons
such as
fundamental or cost-imposed
measurement limits,
data storage constraints,
and
privacy.
Text corpora generate especially heavy-tailed distributions,
with hapax legomena taking up roughly half of a text's lexicon~\cite{simon1955a}.
The Google Books $n$-gram corpus only includes $n$-grams
which have appeared 40 or more times~\cite{michel2011a},
excluding a vast number of rare $n$-grams.
In our present work, we have already seen that for Twitter, our sample is approximately 10\% of all tweets
(with Twitter itself being a rather small subsample of all forms of human expression),
and that baby names with counts of 4 or less are not made public for
any censused population within the US.
Limits to sampling in ecological systems can be severe---the Barro Colorado Island data
is evidently not inclusive of all plant matter.

To investigate the problem of truncation, we explore our
four case studies of
Twitter, tree species, names, and companies
by systematically
limiting the observable components of each system.
For each pair of systems, we take the top $N$=$10^k$ ranked components where
$k$=1.5, 2.0, 2.5, \ldots, stopping once we exceed
the size of both systems.
For each $k$, we generate
the corresponding series of rank-turbulence divergence graphs,
producing
Flipbooks~\flipbooktwittertrunc--\flipbookcompaniestrunc.
For a visual summary of these Flipbooks, we put together
a subset of the rank-rank histograms
to form
Fig.~\ref{fig:rankturbdiv.truncation}.

The five rows of Fig.~\ref{fig:rankturbdiv.truncation}
correspond to our four case studies, with baby names contributing two rows.
The first two examples of Twitter and tree species show
a regular trend towards the full histogram.
By contrast, baby names and market caps both appear to
be disjoint when strong truncation is applied (small $N$).
As $N$ increases, the internal random structure for baby names
and the stable vertical structure for market caps start to be revealed
by $N$=1,000.

For the Flipbooks, we use the same values of $\alpha$
for Twitter $\alpha$=1/3,
tree species $\alpha$=0,
and
market caps $\alpha$=1/3
(Figs.~\ref{fig:rankturbdiv.allotaxonometer9000-2016-11-09-2017-08-13-story-wrangler-all-rank-div_05},
\ref{fig:rankturbdiv.allotaxonometer9000-1985-2015-barro-colorado002},
and
\ref{fig:rankturbdiv.allotaxonometer9000-siblis_2007-Q4_2018-Q4_marketcaps004}).
For baby names, we take the $\alpha=0$ limit as this is the most challenging for the
truncated version.

In general, as $N$ is increased, we see the main stories and patterns emerge.
For Twitter, the election's imprint is clear for low $N$ (Flipbook~\flipbooktwittertrunc)
with the texture of Charlottesville requiring more words to be included.
The most dramatic changes in the lists of rank-turbulence divergence occur
for baby names and market caps,
as the system exclusive types of these comparisons are masked for low $N$.

As a rough rule of thumb, the appearance of separated system-exclusive lines
suggests that the underlying datasets are sufficiently rich enough
to allow for a substantive allotaxonometric comparison.
For the example of Twitter,
and understanding that cell size matters,
we see the separation occurs
when $N$ is moved from 100,000 to 1,000,000.
We see no such separation for tree species however the vertical form
representing stability unveils itself with increasing $N$  in clear fashion.

\section{Guide to Flipbooks}
\label{sec:rankturbdiv.flipbooks}

To help demonstrate rank-turbulence divergence as an allotaxonometric instrument,
we have referenced a number of Flipbooks throughout the paper.
We include these and other Flipbooks as supplementary information
which can be found as part of our paper's online appendices
at
\href{http://compstorylab.org/allotaxonometry/flipbooks}{http://compstorylab.org/allotaxonometry/flipbooks}.

Flipbooks are best `flipped through' back and forth using a PDF reader
with the view set to `single page' rather than continuous.

We list and briefly describe all Flipbooks here.
Our flipbooks follow various formats
which include:
Comparisons of two systems with varying
rank-turbulence divergence parameter $\alpha$;
Comparisons of a series of system pairs, often through time;
and
Comparisons of systems with truncation applied (Sec.~\ref{subsec:rankturbdiv.truncation}).

When $\alpha$ is varied the values
are
$0$,
$\frac{1}{12}$,
$\frac{2}{12}$,
$\frac{3}{12}$,
$\frac{4}{12}$,
$\frac{5}{12}$,
$\frac{6}{12}$,
$\frac{8}{12}$,
$1$,
$2$,
$5$,
and
$\infty$.

\medskip
\textbf{Flipbook \flipbooktwitter---Word use on Twitter:}
US Presidential Election (2016-11-09)
versus the Charlottesville Unite the Right Rally (2017-08-13);
Variation of $\alpha$.

\textbf{Flipbook \flipbooktwitterRT---Word use on Twitter:}
US Presidential Election (2016-11-09)
versus the Charlottesville Unite the Right Rally (2017-08-13);
Variation of inclusion of retweets from 1\% to 100\%;
$\alpha = 1/3$.

\textbf{Flipbook \flipbooktwittertimediff---Word use on Twitter:}
Variation of time comparing 2019/01/04 going forward
roughly logarithmically in number of days to a year ahead,
2020/01/03, the day of the assassination of Qasem Soleimani;
$\alpha = 1/3$.

\smallskip
\textbf{Flipbook \flipbooktrees---Tree species abundance on Barro Colorado Island:}
Fig.~\ref{fig:rankturbdiv.allotaxonometer9000-1985-2015-barro-colorado002}
with variation of $\alpha$.
The Flipbook shows how increasing $\alpha$ from 0 leads to an increasingly poor fit on
the rank-rank histogram.

\smallskip
\textbf{Flipbook \flipbookgirlsyears---Baby girl names over time:}
Described in Sec.~\ref{subsec:rankturbdiv.alphainfty},
comparisons of baby girl name distributions 50 years apart
starting in 1880 and going forward in 5 year increments,
with $\alpha = 1/3$.
Ends with Fig.~\ref{fig:rankturbdiv.allotaxonometer9000-babynames-girls}.

\smallskip
\textbf{Flipbook \flipbookgirlsalphas---Baby boy names over time:}
Described in Sec.~\ref{subsec:rankturbdiv.alphainfty},
comparisons of baby girl name distributions 50 years apart
starting in 1880 and going forward in 5 year increments,
with $\alpha = 1/3$.
Ends with Fig.~\ref{fig:rankturbdiv.allotaxonometer9000-babynames-boys}.

\smallskip
\textbf{Flipbook \flipbookboysyears---Baby girl names, 1968--2018:}
Described in Sec.~\ref{subsec:rankturbdiv.alphainfty},
shows effect of varying $\alpha$,
with Fig.~\ref{fig:rankturbdiv.allotaxonometer9000-babynames-girls}
as the fifth page.

\smallskip
\textbf{Flipbook \flipbookboysalphas---Baby boy names, 1968--2018:}
Described in Sec.~\ref{subsec:rankturbdiv.alphainfty},
shows effect of varying $\alpha$,
with Fig.~\ref{fig:rankturbdiv.allotaxonometer9000-babynames-boys}
as the fifth page.

\smallskip
\textbf{Flipbook \flipbookmarketcapsyears---Market caps:}
Comparison of market caps for publicly traded companies
in the fourth quarter six years apart,
starting with 1995 versus 2001 and ending with 2012 versus 2018,
and with $\alpha$ fixed at 1/3.

\smallskip
\textbf{Flipbook \flipbooktwittertrunc---Word use on Twitter, truncated:}
Full series of allotaxonographs 
corresponding to histograms of
row 1 in Fig.~\ref{fig:rankturbdiv.truncation} with $\alpha=1/3$.

\smallskip
\textbf{Flipbook \flipbooktreestrunc---Tree species abundance, truncated:}
Full series of allotaxonographs 
corresponding to histograms of
row 2 in Fig.~\ref{fig:rankturbdiv.truncation} with $\alpha=0$.

\smallskip
\textbf{Flipbook \flipbookgirlnamestrunc---Baby girl names, truncated:}
Full series of allotaxonographs 
corresponding to histograms of
row 3 in Fig.~\ref{fig:rankturbdiv.truncation} with $\alpha=\infty$.

\smallskip
\textbf{Flipbook \flipbookboynamestrunc---Baby boy names, truncated:}
Full series of allotaxonographs 
corresponding to histograms of
row 4 in Fig.~\ref{fig:rankturbdiv.truncation} with $\alpha=\infty$.

\smallskip
\textbf{Flipbook \flipbookcompaniestrunc---Market caps, truncated:}
Full series of allotaxonographs 
corresponding to histograms of
row 5 in Fig.~\ref{fig:rankturbdiv.truncation} with $\alpha=1/3$.

\smallskip
\textbf{Flipbook \flipbooknba---Season total points scored by players in the National Basketball Association:}
Season to season comparison of total player points per season, $\alpha$ = 1/3.
The Flipbook starts with 1996--1997 versus 1997--1998
and ends in
2017--2018 versus 2018--2019.
Rookies, retirements, injuries are all in evidence.
For $\alpha=1/3$, Carmelo Anthony in 2003--2004 has the strongest debut, just ahead
of Lebron James in the same year.
Overall, Dwyane Wade's 2008--2009 season produced
the highest $\rtdelement{1/3}$, moving from
$\zipfrank$=51 to 1 over the previous year where
he was limited in playing time with injuries.
In 2008--2009, Wade's points per game of 30.2 would be the highest of his career
but his team, the Miami Heat, would founder, achieving the worst record in the NBA.

\smallskip
\textbf{Flipbook \flipbookgoogleonegrams---Google Books, Fiction in 1948 versus 1987, 1-grams:}
The first of three Flipbooks exploring $n$-gram usage in books
by varying $\alpha$.
We have elsewhere documented the deeply problematic influence of
scientific literature and individual books in Ref.~\cite{pechenick2015a},
rendering the Google Books project unreliable, as is.
Nevertheless, the Version 2 $n$-grams dataset for English fiction
is worth exploring~\cite{pechenick2017a}
with different instruments,
and we are endeavoring separately to provide corrective measures.
For 1948, we see characters and place names dominate, and these
come from a few books (e.g., `Lanny Budd', `Raintree County').
The 1987 side shows words that are not tied to specific books
but rather cultural and temporal phenomena, as well as cruder language:
`KGB', `CIA', `Vietnam',
`lesbian', `television', `computer', and `fucking'.
Tuning $\alpha$ towards $\infty$, we can see pronouns changing slightly in rank
with `her and `she' elevating and `he' and `his' dropping.

\smallskip
\textbf{Flipbook \flipbookgooglebigrams---Google Books, Fiction in 1948 versus 1987, 2-grams:}
For 2-grams, we again see character names dominate 1947 for low $\alpha$
(`Sung Chiang', `the Perfessor'), while `the CIA' and `the KGB' stand out for 1987.
Increasing $\alpha$ brings in the same words as for 1-grams preceded by `the'
(`the phone', `the computer').
As $\alpha \rightarrow \infty$, bigrams with `not' as part appear more strongly for 1987.

\smallskip
\textbf{Flipbook \flipbookgoogletrigrams---Google Books, Fiction in 1948 versus 1987, 3-grams:}
For 3-grams, while we still see characters and place names for 1947, we now
have what we call `pathological hapax legomena', words (or trigrams in this case) that occur
once in many books.  The 3-grams are all from standardized, legal-speak front matter
coming from outside of the story: `change without notice', `your local bookstore', and `Cover art by'.
A second kind of trigram that dominates appears to be one that appears as
part of a book's title printed on every page in the header or footer.
As we increase $\alpha$, we again see 'not' appearing in contributing 1987 trigrams.
Because of the combinatorial explosion around words like `computer' and `phone',
we no longer see them in the trigram lists.
One upshot of this brief inspection of Google Books is to highlight the value of
separately examining $n$-grams.
We also note that the 3-gram example is our largest system-system comparison with
system sizes on the order of $10^9$.

\smallskip
\textbf{Flipbook \flipbookharrypotter---Harry Potter books, all 1-grams:}
Comparison of each Harry Potter book relative to all all other books in the series
combined, using $\alpha$=1/2
(the single book is the right hand system, the merged set of 6 books the left system).
Character names and major objects and places dominate,
and the first book is most different from the others combined.

\smallskip
\textbf{Flipbook \flipbookharrypotternocaps---Harry Potter books, uncapitalized 1-grams:}
The same comparison as the previous Flipbook but now with
all capitalized words excluded, as an example attempt to
use a different lens on our allotaxonometer.
Hagrid's speech in part separates Book 1 ('yer', `ter'),
Book 3 has `rat', `dementor', and a relative abundance of em dashes ('---'),
Book 7 has `sword', `wand', and `goblin'.
The dominant elements are things, places, and repeated
actions (e.g., spells) and descriptors.
To examine changes in functional word usage, which may reveal
changes in Rowling's writing, we would increase $\alpha$
as we did for Google Books.
Again, we see the relative ease of taking subsets with ranks
for allotaxonometry.

\smallskip
\textbf{Flipbook \flipbookdeathcauses---Causes of Death in Hong Kong:}
Five year gap comparison of causes of death reported per year in Hong Kong,
starting with 2001 versus 2006 and moving through
to 2012 versus 2017.
Overall, pneumonia is the leading cause of death.
In the second half of the time frame,
`kidney disease' and `dementia'
stand out as becoming more prevalent.
Deaths listed as due to heroin drop off markedly in 2012 and 2013
relative to 5 years before.
We note that changes in diagnoses, practices, and
categorization are all confounding issues.

\smallskip
\textbf{Flipbook \flipbookjobnames---Job titles:}
US job titles based on text analysis of online postings,
2007 compared with 2018;
variation across three kinds of job categorization,
from coarse- to fine-grained groupings,
with suitable variation of $\alpha$
($\alpha=0$, $\alpha=1/12$, and $\alpha=1/3$).

\section{Data and Code}
\label{sec:rankturbdiv.methods}

\subsection{Datasets}
\label{sec:rankturbdiv.methods.datasets}

\textbf{Word usage on Twitter:}
Derived from an approximate 10\% sample of Twitter
collective by the Computational Story Lab from 2008 to 2020;
English language detection performed per Ref.~\cite{alshaabi2020a}.

\textbf{Species abundance on Barro Colorado Island:}
The dataset and its online repository for censuses taken over 35 years
are described in Ref.~\cite{condit2019a}.

\textbf{Baby names:}
Data taken from Social Security Card applications.
For each year from 1880--2018, the dataset includes all names which have 5 or more applications.
Because Social Security Numbers were first issued at the end of 1936, there is a change in
the dataset's nature as people moved from
registering as adults to being solely registered at birth.
While we use the dataset as is here, we note that there is a clear change in the
male to female ratio with more boys being registered from 1940 onwards.
Baby name dataset available here:\\
\href{https://catalog.data.gov/dataset?tags=baby-names}{https://catalog.data.gov/dataset?tags=baby-names}.
Separate dataset for total births available here:\\
\href{https://www.ssa.gov/oact/babynames/numberUSbirths.html}{https://ssa.gov/oact/babynames/numberUSbirths.html}.

\textbf{Market cap data:}
The underlying dataset comprises 9,322 US publicly traded companies that have been
part of the S\&P 500 at any point during the period of 1979--2018,
or part of the Russell 3000 index from 1995 on.
Data is available from Siblis Research here:
\href{http://siblisresearch.com/data/us-equity-returns/}{http://siblisresearch.com/data/us-equity-returns/}.

\textbf{National Basketball Association:}
Dataset available here:
\href{https://stats.nba.com/players/traditional/}{https://stats.nba.com/players/traditional/}.

\textbf{Google Books $n$-grams:}
Version 2, English Fiction.
We filtered the database to
collect only $n$-grams containing simple latin characters.
Dataset available here~\cite{michel2011a}:
\href{https://books.google.com/ngrams}{https://books.google.com/ngrams}.

\textbf{Causes of Death in Hong Kong}
The dataset is described in
Ref.~\cite{hongkong_dataset2018a,hongkong_districts2016a,tertiary-planning-units2016a}
and has been well studied by others~\cite{wong2001a,lam2004a,ou2008a,qiu2015a,wong2015a,wu2017a}.
The dataset contains 892,055 death records between 1995 and 2017.
  
\textbf{Job titles:}
Provided by Burning Glass, the dataset is derived from
online postings (several million job openings per day, tens of thousands of sources).
Raw listings are processed and categorized into two smaller
taxonomies with natural-language algorithms.

\subsection{Code}
\label{sec:rankturbdiv.code}

All scripts and documentation reside on Gitlab:
\href{https://gitlab.com/compstorylab/allotaxonometer}{https://gitlab.com/compstorylab/allotaxonometer}.

For the present paper, we wrote the scripts to generate
the allotaxonographs in MATLAB (Laboratory of the Matrix).
We produced all figures and flipbooks using MATLAB Versions R2019b and R2020a.
The core script is highly configurable
and can be used to create a range of allotaxonographs
as well as simple unlabeled rank-rank histograms.
Instruments accommodated by the script include
rank-turbulence divergence,
probability-turbulence divergence~\cite{dodds2020g},
and generalized symmetric entropy divergence which
includes Jensen-Shannon divergence as a special case.

\section{Concluding remarks}
\label{sec:rankturbdiv.concludingremarks}

\begin{table*}
  \begin{tabular}{|c|c|c|c|}
    \hline
    \hline
    Systems
    $\systema$
    and
    $\systemb$
    &
    Visualization/Section
    &
    $\alpha$
    &
    $\rtdalphasystemsOmega$
    \\
    \hline
    \hline
    Matching systems
    &
    Fig.~\ref{fig:rankturbdiv.zipfturbulence-2016-11-09-2017-08-13-story-wrangler-all-rank-bare}B,
    Sec.~\ref{subsec:rankturbdiv.viz}
    &
    Any
    &
    0
    \\
    \hline
    Species, Barro Colorado Island
    &
    Fig.~\ref{fig:rankturbdiv.allotaxonometer9000-1985-2015-barro-colorado002},
    Sec.~\ref{subsec:rankturbdiv.alpha0}
    &
    0
    &
    0.077
    \\
    \hline
    Causes of death, Hong Kong, 2012--2017
    &
    Flipbook~\flipbookdeathcauses,
    Sec.~\ref{sec:rankturbdiv.flipbooks}
    &
    1/3
    &
    0.213
    \\
    \hline
    Player season points, 2014--2015 vs 2015--2016
    &
    Flipbook~\flipbooknba,
    Sec.~\ref{sec:rankturbdiv.flipbooks}
    &
    1/3
    &
    0.279
    \\
    \hline
    Lowercase words, Harry Potter 7 vs vs 1--6
    &
    Flipbook~\flipbookharrypotternocaps,
    Sec.~\ref{sec:rankturbdiv.flipbooks}
    &
    1/2
    &
    0.308
    \\
    \hline
    Companies, 2007 vs 2018
    &
    Fig.~\ref{fig:rankturbdiv.allotaxonometer9000-siblis_2007-Q4_2018-Q4_marketcaps004},
    Sec.~\ref{subsec:rankturbdiv.shocks}
    &
    1/3
    &
    0.441
    \\
    \hline
    Words on Twitter:
    2016/11/09 vs 2017/08/13
    &
    Fig.~\ref{fig:rankturbdiv.allotaxonometer9000-2016-11-09-2017-08-13-story-wrangler-all-rank-div_05},
    Sec.~\ref{subsec:rankturbdiv.construction}
    &
    1/3
    &
    0.493
    \\
    \hline
    Baby boy names, US, 1885 vs 1910
    &
    Sec.~\ref{subsec:rankturbdiv.alphainfty}
    &
    $\infty$
    &
    0.536
    \\
    \hline
    Baby girl names, US, 1900 vs 1925
    &
    Sec.~\ref{subsec:rankturbdiv.alphainfty}
    &
    $\infty$
    &
    0.631
    \\
    \hline
    Baby boy names, US, 1918 vs 1968
    &
    Flipbook~\flipbookboysyears,
    Sec.~\ref{subsec:rankturbdiv.alphainfty}
    &
    $\infty$
    &
    0.772
    \\
    \hline
    Baby boy names, US, 1968 vs 2018
    &
    Fig.~\ref{fig:rankturbdiv.allotaxonometer9000-babynames-boys},
    Sec.~\ref{subsec:rankturbdiv.alphainfty}
    &
    $\infty$
    &
    0.850
    \\
    \hline
    Baby girl names, US, 1918 vs 1968
    &
    Flipbook~\flipbookgirlsyears,
    Sec.~\ref{subsec:rankturbdiv.alphainfty}
    &
    $\infty$
    &
    0.887
    \\
    \hline
    Baby girl names, US, 1968 vs 2018
    &
    Fig.~\ref{fig:rankturbdiv.allotaxonometer9000-babynames-girls},
    Sec.~\ref{subsec:rankturbdiv.alphainfty}
    &
    $\infty$
    &
    0.926
    \\
    \hline
    Disjoint systems
    &
    Fig.~\ref{fig:rankturbdiv.zipfturbulence-2016-11-09-2017-08-13-story-wrangler-all-rank-bare}D,
    Sec.~\ref{subsec:rankturbdiv.viz}
    &
    Any
    &
    1
    \\
    \hline
    \hline
  \end{tabular}
  \caption{
    A selection of example system comparisons
    producing 
    a range of $\rtdalphasystemsOmega$ values.
  }
  \label{fig:rankturbdiv.RTDexamples}
\end{table*}

Our goal has been to propose, advocate
for, and contribute to a field of allotaxonometry:
The measurement and visualization of
detailed, type-level differences between complex systems.
In the development of dynamic
allotaxonometric dashboards,
we have argued for a full embrace of complexity
and stringent avoidance
of falling into the trap of describing system differences
solely by a single number.

In Sec.~\ref{subsec:rankturbdiv.introduction-motivation},
we observed numerous benefits for using ranks:
Widespread applicability beyond systems with type frequencies, probabilities, or rates;
a natural handling of system exclusive types by ranking them last;
robustness of rank-based statistics,
and
the straightforward interpretability of ranked lists.

Focusing on systems with many components which
can be ranked by some kind of well-defined size,
we have created, tested, and explored
rank-based allotaxonographs
built around our conception of a tunable rank-turbulence divergence.
In Tab.~\ref{fig:rankturbdiv.RTDexamples},
we collect a list of example system comparisons
with $\rtdalphasystemsOmega$ ranging from 0 to 1.

At the core of rank-turbulence divergence
in
\Req{eq:rankturbdiv.rankturbdiv_good}
is the
interpretable difference of inverse powers of type ranks:
\begin{equation}
\left\lvert
\frac{1}{\left[\zipfrank_{\elementsymbol,\indexaraw}\right]^{\alpha}}
-
\frac{1}{\left[\zipfrank_{\elementsymbol,\indexbraw}\right]^{\alpha}}
\right\rvert.
\label{eq:rankturbdiv.core}
\end{equation}
As $\alpha \rightarrow 0$, the differences between ranks are contracted
and low rank types become more salient.
As $\alpha \rightarrow \infty$, rank discrepancies
become more exacerbated, and the highest rank types dominate.

Narrowing our view to systems which afford frequencies of components,
we find our directly tunable divergence appears to be far
more general
than many probability-based divergences,
which are largely grouped around a few core structures.
Per~\cite{cha2007a} and imposing the Zipfian ideal of
$p = 1/\zipfrank$, we see that
$
\left|
\zipfrank_{\elementsymbol,\indexaraw}^{-1}
-
\zipfrank_{\elementsymbol,\indexbraw}^{-1}
\right|
$
is an abundant form.
There are a few other variations including
$
\min
\left(
\zipfrank_{\elementsymbol,\indexaraw},
\zipfrank_{\elementsymbol,\indexbraw}
\right),
$
and the Hellinger-like distance
$
\left|
\zipfrank_{\elementsymbol,\indexaraw}^{-\onehalf}
-
\zipfrank_{\elementsymbol,\indexbraw}^{-\onehalf}
\right|.
$
These three cases correspond to
our rank-turbulence divergence
with $\alpha$=1, $\infty$, and $1/2$.

For the instrument's integrity and power,
we assert that the map and list should be bound together.
While our allotaxonomic histograms give immediate stories from the
automatically labeled words along the fringes, the overall ordering
of these words by some measure of importance is unclear.
And in choosing to map a two-dimensional rank-rank histogram
onto a single dimension---another ranked list---we remain mindful that we
are discarding information.
We suggest that, analogously, all cartograms would benefit from an associated
ordered list and vice versa~\cite{alajajian2017a}.

Per our introduction, there is tendency across diverse fields
towards creating single-number measurements of complex systems,
and that this is especially problematic when
heavy-tailed Zipf distributions are in evidence.
We have shown that even when single-number measures match
for two systems, allotaxonographs using rank-turbulence divergence
are able to reveal and make sense of the full variation
between systems.

The four main case studies of
Twitter,
tree species,
baby names,
and
companies
all provided rich and diverse examples of allotaxonometric comparisons.
Our ability to readily analyze the effects of partially sampled data
in Sec.~\ref{subsec:rankturbdiv.truncation} further showed
the value of a rank-based approach.

There are many future research possibilities, both theoretical and applied,
suggested or opened up by what we have developed here for rank-turbulence divergence
and, more generally, for allotaxonometry.

With our supplementary Flipbooks,
we have attempted to show the
prospect for the building of online, interactive allotaxonographs.
Being linear in nature, Flipbooks allow us to explore one dimension
of variation at a time, and by design are built to be fixed rather than flexible.
For baby names, for example, we would like to 
be able to interactively vary the years being compared as well as
rank-turbulence divergence's $\alpha$.
For temporally evolving systems,
an interactive allotaxonograph could be
set to track a particular cohort of types or
to automatically highlight those
which make a dynamical transition of some prescribed kind.

We have been pragmatic in our construction of rank-turbulence divergence,
striving to build a functional tool first and foremost.
A rigorous theoretical foundation might be possible for
either our tool or an adjacent rank-based divergence.
Staying on the functional side, variations on our divergence might be
of use for some comparisons where no value of $\alpha$ makes for a
good fit.
As we noted for the case of market caps,
a composite instrument that separates stable, enduring companies
to those that exit or enter could be devised.

For systems with documented component probabilities or rates,
we have also constructed a related probability-turbulence divergence.
We explore the allotaxonometry of this divergence in~\cite{dodds2020g},
showing the instrument to be a generalization of a suite of
well known probability-based divergences.

When rank turbulence is in evidence, as in the case of Twitter,
we would want to be able to determine an optimal $\alpha$.
While for generalized entropy approaches for single systems, the limit of
linear scaling and Shannon's entropy demarcate the boundary
between accentuating the common or the rare~\cite{hill1973a,tsallis2001a,keylock2005a,jost2006a},
we have found that for system comparisons, the optimal value of $\alpha$, if it exists,
is dependent on the pair of systems being compared.

In the present work, we have left open the
possibility of an analytic
connection between the rank-turbulence scaling described
at the end of Sec.~\ref{subsec:rankturbdiv.introduction-rankings},
and, to the extent that well-defined scaling is present,
with an optimal $\alpha$ for rank-turbulence divergence.

For another direction, we venture that a kind of `rank energy' interpretation might be possible.
Working from the idealized Zipf relationship of $p \sim r^{-1}$, we would have
\begin{equation}
  p^{\alpha}
  \sim
  1/\zipfrank^{\alpha}
  =
  \textnormal{exp}
  \left\{
  -\alpha E/T
  \right\}
  =
  \textnormal{exp}
  \left\{
  -E/T'
  \right\},
  \label{eq:rankturbdiv.rankenergy}
\end{equation}
where $E = T \ln \zipfrank$
is an energy associated with rank $\zipfrank$
and temperature $T$,
and $T'$ an effective temperature.
When $T' \rightarrow 0$, high ranked types prevail,
while when
$T' \rightarrow \infty$,
all types move towards being weighted equally, independent of rank.

As we saw for the unusually durable popular name `Elizabeth'
in Fig.~\ref{fig:rankturbdiv.allotaxonometer9000-babynames-girls},
there are components whose locations on allotaxonographs are not highlighted
by standard conceptions of divergences, rank-based or otherwise.
A completely distinct measure of importance could favor
largely isolated rank-rank pairings
on the rank-rank histogram.
Given that the measure would have to be sufficiently sophisticated
to accommodate the possibility that a small
cluster of related types might be near each other (e.g., `Lady' and `Gaga'),
yet otherwise be distinct, the application of
some basic kind of cluster analysis would offer a starting point.

We close with the observation that
in terms of applications, any comparison of complex systems entailing
a broad array of components would be fair game.
A few examples would be sales of anything (e.g., Amazon's sales from week to week),
crime rates,
country exports,
sites visited or searched for online,
medical condition prevalences,
rankings in sports,
music popularity,
and
markets of all kinds.
And while our focus has been on comparing systems at the level of components,
changes in system structure, e.g., complex networks, could also
be readily
explored with the same rank-turbulence divergence instrument.

\acknowledgments
The authors are grateful for 
support furnished by MassMutual and Google,
and the computational facilities provided
by the Vermont Advanced Computing Core.
The authors are grateful for converstations
with R. Gallagher, L. Mitchell, and J. Weitz.

\clearpage

\end{document}